\documentclass[twocolumn,preprint]{aastex631}

\usepackage[utf8]{inputenc}
\usepackage{CJK}
\usepackage[utf8]{inputenc}
\usepackage[flushleft]{threeparttable}
\usepackage{longtable}
\usepackage{multirow}
\usepackage[T1]{fontenc}
\usepackage{bm}
\usepackage{enumitem}
\usepackage{array}
\usepackage{mathtools}
\makeatletter

\newcommand{\Rmnum}[1]{\expandafter\@slowromancap\romannumeral #1@}
\makeatother

\shorttitle{Bridging the surface and core of RSG progenitor}
\shortauthors{Fang et al.}

\begin{document}

\title{Diversity in Hydrogen-rich Envelope Mass of Type II Supernovae. (III). The mass-loss and evolutionary pathways of the red supergiant progenitors}
\author[0000-0002-1161-9592]{Qiliang Fang}\affiliation{National Astronomical Observatory of Japan, National Institutes of Natural Sciences, 2-21-1 Osawa, Mitaka, Tokyo 181-8588, Japan}
\author[0000-0003-1169-1954]{Takashi J. Moriya}
\affiliation{National Astronomical Observatory of Japan, National Institutes of Natural Sciences, 2-21-1 Osawa, Mitaka, Tokyo 181-8588, Japan}
\affiliation{Graduate Institute for Advanced Studies, SOKENDAI, 2-21-1 Osawa, Mitaka, Tokyo 181-8588, Japan}
\affiliation{School of Physics and Astronomy, Monash University, Clayton, Victoria 3800, Australia}
\author[0000-0003-2611-7269]{Keiichi Maeda}\affiliation{Department of Astronomy, Kyoto University, Kitashirakawa-Oiwake-cho, Sakyo-ku, Kyoto 606-8502, Japan}
\author{Andris Dorozsmai}\affiliation{National Astronomical Observatory of Japan, National Institutes of Natural Sciences, 2-21-1 Osawa, Mitaka, Tokyo 181-8588, Japan}
\author[0009-0000-2457-279X]{Javier Silva-Farfán}\affiliation{Department of Astronomy, Kyoto University, Kitashirakawa-Oiwake-cho, Sakyo-ku, Kyoto 606-8502, Japan}

\begin{abstract}
We present a comprehensive analysis of 32 type II supernovae (SNe II) with plateau phase photometry and late phase ($nebular$) spectroscopy available, aiming to bridge the gap between the surface and core of their red supergiant (RSG) progenitors. Using \texttt{MESA}\,+\texttt{STELLA}, we compute an extensive grid of SN II light curve models originating from RSG with effective temperatures $T_{\rm eff}$ around 3650\,K and hydrogen-rich envelopes artificially stripped to varying degrees. These models are then used to derive the hydrogen-rich envelope masses $M_{\rm Henv}$ for SNe II from their plateau phase light curves. Nebular spectroscopy further constrains the progenitor RSG's luminosity log\,$L_{\rm prog}$, and is employed to remove the degeneracies in light curve modeling. The comparison between log\,$L_{\rm prog}$-$M_{\rm Henv}$ reveals that $M_{\rm Henv}$ spans a broad range at the same log\,$L_{\rm prog}$, and almost all SNe II have lower $M_{\rm Henv}$ than the prediction of the default stellar wind models. We explore alternative wind prescriptions, binary evolution models, and the possibility of more compact RSG progenitors. Although binary interaction offers a compelling explanation for the non-monotonicity and large scatter in the log\,$L_{\rm prog}$-$M_{\rm Henv}$ relation, the high occurrence rate of partially-stripped RSGs cannot be accounted for by stable binary mass transfer alone without fine-tuned orbital parameters. This highlights that, despite being the most commonly observed class of core-collapse SNe, SNe II likely originate from a variety of mass-loss histories and evolutionary pathways that are more diverse and complex than typically assumed in standard stellar evolution models. %

\end{abstract}
 


\section{INTRODUCTION}
Type II supernovae (SNe II) are believed to originate from the iron-core infall of massive star, with zero-age-main-sequence (ZAMS) mass $M_{\rm ZAMS}\,$\(>\)\,8\,$M_{\rm \odot}$ at the end of their lives, while still retaining a relatively massive hydrogen-rich envelope. Direct imaging of the progenitors of SNe II suggests that most of them originate from the explosion of red supergiants (RSGs; see \citealt{smartt15} for a review), as predicted by modern stellar evolution models. Being one of the most common type of core-collapse supernovae (CCSNe; \citealt{li06,perley20,sharma24}), SNe II play a crucial role in understanding the core-collapse process and the evolutionary pathways that lead to the formation of their progenitors. For these purposes, it is important to infer the physical properties of SNe II from observations, such as the zero-age-main-sequence (ZAMS) mass ($M_{\rm ZAMS}$) of the progenitors, the explosion energy ($E_{\rm K}$) and the hydrogen-rich envelope mass ($M_{\rm Henv}$), and to explore their ranges as well as potential correlations.

One of the commonly adopted method to constrain these quantities is to model the light curves of SNe II. Following the expansion, the expelled material (ejecta) cools down, and the hydrogen elements recombine to form a sharp ionization front, causing the ejecta to radiate at a nearly-constant luminosity for $\sim$\,100 days, known as the plateau phase \citep{kasen09}. The duration and the magnitudes during this phase are affected by many factors, including the hydrogen-rich envelope mass $M_{\rm Henv}$, the radius of the progenitor $R_{\rm prog}$, the explosion energy $E_{\rm K}$ and the mass of the radioactive $^{56}$Ni deposited in the ejecta $M_{\rm Ni}$ (\citealt{kasen09,dessart19,goldberg19,fang25a}). These parameters can be inferred by fitting observed light curves with model light curves from progenitor models with varying pre-explosion structures, explosion energies, and $^{56}$Ni masses (\citealt{morozova18,martinez22a,martinez22,martinez22c,moriya23}).

The pre-SN structure can be constructed in a parameterized way, or from evolutionary models. In the latter case, in general, a non-rotating star is evolved from ZAMS phase to the onset of core-collapse, with a selected wind mass-loss scheme. However, this simple picture can be complicated by several factors:  (1) The wind-driven mass-loss rate, which depends on the properties of the progenitor, is quite uncertain from both observational (\citealt{beasor20,wang21,yang23,antoniadis24,decin24,zapartas24} for recent works) and theoretical sides (\citealt{kee21,vink23,cheng24,fuller24}). Stellar evolution models usually adopt the empirical mass-loss prescriptions (\citealt{dejager88,nieu90,vanbeveren98,vanloon05,goldman17,beasor20,wang21}) which may vary by an order of magnitude (see comparison of these mass-loss prescriptions in \citealt{mauron11} and \citealt{yang23}); (2) Massive stars often form in binary systems, where interactions can efficiently strip the hydrogen envelope to varying degrees (\citealt{heger03, eldridge08, yoon10, smith11, sana12, smith14, yoon15, yoon17, ouchi17, eldridge18, fang19, zapartas19, laplace20, renzo21, zapartas21,klencki22,chen23, ercolino23, fragos23, hirai23,sun23,fang24,matsuoka23}). These factors introduce significant scatter in the pre-SN structure, even among progenitors with the same $M_{\rm ZAMS}$ (see e.g. \citealt{renzo17}), making the inference of $M_{\rm ZAMS}$ from light curve modeling uncertain. Growing evidence suggests that enhanced mass-loss is required to explain the light curves of several individual events \citep{hiramatsu21, 2020jfo, 2018gj, 2021wvw, hsu24, fang25b}. Furthermore, \citet{fang25a} propose that nearly half of SN II progenitors have undergone partial removal of their hydrogen-rich envelopes to $M_{\rm Henv}$\,\(<\)\,5.0\,$M_{\rm \odot}$ prior to explosion, accounting for the observed diversity in SN II light curves (see, e.g., \citealt{anderson14,valenti16,anderson24}).

Although the surface properties of the RSG progenitors are sensitive to the mass-loss mechanism, the nucleosynthesis products within the helium core are less affected by the amount of the hydrogen-rich envelope (e.g., \citealt{sukhbold18,takahashi23})\footnote{However, recent studies indicate systematic differences in the core structures of single stars and binary-stripped stars if the core still develops after a large amount of hydrogen-rich envelope is removed \citep{laplace21, schneider21, farmer21, farmer23}.}. Several months to a year after the explosion, the ejecta becomes transparent and enters $nebular$ phase, during which the spectroscopy is dominated by emission lines from intermediate-mass elements. In particular, the relative strength of the oxygen emission [O I] $\lambda\lambda$6300,6363 is sensitive to the mass of the oxygen elements $M_{\rm O}$ in the ejecta, and is considered as a reliable tracer of $M_{\rm ZAMS}$ if the $M_{\rm ZAMS}$-$M_{\rm O}$ relation is assumed (\citealt{fransson89,maeda07,jerkstrand12,jerkstrand14,hanin15,jerkstrand15,jerkstrand17,fang18,fang19,prentice19,terreran19,dessart20,dessart21a,dessart21b,hiramatsu21,fang22,dessart23,fang25b}). Independent from plateau phase light curve modeling, nebular spectroscopy offers additional insights into the properties of SNe II. 

The purpose of this work is to bridge the gap between light curve modeling, which reflects the outer properties of the progenitor, and nebular spectroscopy, which probes the core. Given the large uncertainties in the mass-loss mechanism (driven by stellar wind or binary interaction) and mass-loss rates, in this work, $M_{\rm Henv}$ is treated as a free parameter that can vary by arbitrary degree at a fixed ZAMS mass. We find that for these partially-stripped progenitor models, the key parameters inferred from plateau phase light curve modeling, including $M_{\rm Henv}$, $E_{\rm K}$ and log\,$M_{\rm Ni}$, degenerate with $M_{\rm ZAMS}$. Nebular spectroscopy is employed to break this degeneracy, and to estimate the luminosity of the progenitor RSG log\,$L_{\rm prog}$ based on the empirical mass-luminosity relation (MLR) established in \citet{fang25c}. A comparison between log\,$L_{\rm prog}$ and $M_{\rm Henv}$ for 32 SNe II reveals significant scatter, with nearly all objects showing lower $M_{\rm Henv}$ than predicted by \texttt{KEPLER} models (\citealt{kepler16}). We explore several potential explanations for this discrepancy, including alternative wind prescriptions, binary interactions, and more compact RSG progenitors. Finally, we demonstrate that the initial-mass-function (IMF) incompatibility reported by \citet{martinez22}, characterized by an unexpected large fraction of SNe II with low $M_{\rm ZAMS}$, can be resolved if partially-stripped massive progenitors are considered.

This paper is organized as follows: in \S2, we introduce the numerical set up, including the light curve models, and the progenitor RSGs that are evolved with difference stellar wind prescriptions or in a binary system. In \S3, we describe the fitting procedure used to derive the physical parameters, and explain how nebular spectroscopy is incorporated with light curve modeling. In \S4, we compare the derived log\,$L_{\rm prog}$ and $M_{\rm Henv}$ for the 32 SNe II in our sample and discuss their implications for wind mass-loss rates, binary interactions, and the compactness of RSG progenitor. In \S5, we compare our results with other fitting methods employed in previous studies. Conclusions are left to \S6.

\section{Numerical Setup}
\subsection{Light curve grid}
The light curve models employed in this work are the same as \citet{fang25a,fang25b}, with a denser parameter space. Here we give a brief description on the numerical setup of the light curve models, and readers are encouraged to refer to that paper for further details. 
\begin{itemize}
    \item We first use \texttt{MESA} (version \texttt{r-23.05.01};\citealt{paxton11, paxton13, paxton15, paxton18, paxton19, mesa23}) to initialize a gird of non-rotating progenitors, with solar metallicity $Z$\,=\,0.0142 \citep{nieva12} and $M_{\rm ZAMS}$ ranging from 10 to 18\,$M_{\rm \odot}$ with 1$\,M_{\rm \odot}$ step. The initial setup broadly follows the test suite 20M\_pre\_ms\_to\_core\_collapse, except that the models have mixing length parameter $\alpha_{\rm MLT}$\,=\,2.5, and exponential overshooting parameter $f_{\rm ov}$\,=\,0.004. We specially chose these parameters such that our models have effective temperature $T_{\rm eff}\,\sim\,$3650\,K during RSG phase, consistent with RSGs in the field (\citealt{levesque05}; see also \S4.3), and follow the same $M_{\rm ZAMS}$-$M_{\rm He\,core}$ relation as \texttt{KEPLER} models (\citealt{kepler16}).
    
    These models are first evolved to helium depletion phase, and then the hydrogen-rich envelope is removed to different degree using the command $\texttt{relax\_mass\_to\_remove\_H\_env}$, followed by the evolution until core carbon depletion. Here we do not track the evolution until core-collapse because our primary goal is to model the plateau-phase light curve, which depends almost exclusively on the structure of the hydrogen-rich envelope and is largely decoupled from the advanced burning stages. As demonstrated by \citet{fang25a}, evolving further to core-collapse is computational expensive especially for relatively low-mass progenitors while yielding negligible differences in the plateau-phase light curve (see also \citealt{eldridge19}). In total, the grid contains 9 distinct $M_{\rm ZAMS}$ (or equivalently, log\,$L_{\rm prog}$) values and 118 progenitor models if variations in $M_{\rm Henv}$ are also counted;
    
    \item  The next step is to model the hydrodynamic of the explosion with MESA. The initial setup broadly follows the test suite ccsn\_IIp. The innermost region of the progenitor models are removed, and different amounts of energy are injected in the innermost region as thermal bomb to trigger the explosion. 
    After the shock has reached the surface, we halt the evolution, and deposit different amounts of $^{56}$Ni uniformly below the inner boundary of the hydrogen-rich envelope. We then use the boxcar scheme to smooth the material (\citealt{kasen09,dessart12,dessart13,snec15,fang25a}).

    Here we do not employ the Duffell mixing prescription implemented in MESA \citep{duffell16,paxton18}. As a result, our approach mixes the composition (including $^{56}$Ni), but does not smooth the density profile self-consistently. This distinction may affect the light curve shape, where density mixing, especially of the helium core–envelope interface, has been shown to affect the luminosity drop-off \citep{utrobin07,utrobin17}. Indeed, such effects can be seen in some of our models (e.g., in Figure 1). A systematic exploration of RTI-driven density mixing and its impact on light curves is beyond the scope of this work. All results presented here are based on the simplified boxcar composition mixing as described above.

    The MESA inlists used to generate the progenitor and explosion models are available on Zenodo under an open-source Creative Commons Attribution 4.0 International license: doi:10.5281/zenodo.13953755.
    \item The model is then passed to the public version of $\texttt{STELLA}$, a one-dimensional multi-frequency radiation hydrodynamics code (\citealt{blinnikov98, blinnikov00, blinnikov06}), to model the multi-band light curves. We use 40 frequency bins and 800 spatial zones. Although \citet{paxton18} show that finer frequency resolution is needed to model the multi-band light curve, the difference in the 40-bin and 200-bin setup is marginal and would not significantly affect the overall conclusion in this work. Since the main focus of this work is the plateau phase light curve, no circumstellar medium (CSM) is introduced into the models.
\end{itemize}
The ranges of the parameter are listed in Table~\ref{tab:model_grid}. It should be noted that the intervals of \{$M_{\rm Henv}$, $E_{\rm K}$, $M_{\rm Ni}$\}\footnote{Hereafter, $M_{\rm Henv}$, $E_{\rm K}$ and log\,$M_{\rm Ni}$ are in the units of $M_{\rm \odot}$, foe (10$^{51}$ erg) and dex (with $M_{\rm Ni}$ in the unit of $M_{\rm \odot}$), respectively.} are not uniform. For example, while $E_{\rm K}$ is often sampled at steps of 0.1, in some cases, the interval increases to 0.3.  
Models that failed to converge are discarded, and the final grid consists of 58846 models.

\begin{deluxetable*}{ccccccc|c|c}[t]
\centering
\label{tab:model_grid}
\tablehead{
\colhead{Group number}&\colhead{$M_{\rm ZAMS}$}&\colhead{$M_{\rm He\,core}$}&\colhead{log\,$L_{\rm prog}$}&\colhead{$T_{\rm eff}$}&\colhead{$R_{\rm prog}$}&\colhead{$M_{\rm Henv}$}&\colhead{$E_{\rm K}$}&\colhead{log$\,M_{\rm Ni}$}
}
\startdata
0&10&2.44&4.52&3732$^{+108}_{-114}$&450$^{+28}_{-29}$&[3.0,~7.0]&\multirow{3}{*}{[0.1,~1.5]}&\multirow{9}{*}{[-3,~-0.8]} \\
1&11&2.79&4.62&3677$^{+101}_{-99}$&521$^{+29}_{-28}$&[3.0~\,7.5]&&\\
2&12&3.15&4.72&3644$^{+91}_{-95}$&593$^{+31}_{-32}$&[3.0,~8.0]&&\\
\cline{8-8}
3&13&3.53&4.81&3619$^{+83}_{-83}$&646$^{+29}_{-30}$&[3.0,~9.0]&\multirow{3}{*}{[0.3,~1.5]}&\\
4&14&3.94&4.89&3637$^{+101}_{-98}$&701$^{+33}_{-38}$&[3.0,~10.0]&&\\
5&15&4.36&4.96&3616$^{+93}_{-92}$&769$^{+38}_{-38}$&[3.0,~10.0]&&\\
\cline{8-8}
6&16&4.73&5.02&3602$^{+85}_{-89}$&827$^{+43}_{-38}$&[3.0,~10.0]&\multirow{3}{*}{[0.4,~1.5]}&\\
7&17&5.34&5.09&3588$^{+79}_{-82}$&910$^{+40}_{-35}$&[3.0,~10.0]&&\\
8&18&5.70&5.13&3583$^{+70}_{-71}$&956$^{+40}_{-38}$&[3.0,~10.0]&&
\enddata
\caption{Summary of the model properties. Columns: Group number, ZAMS mass, helium core mass, luminosity of the RSG progenitor, effective temperature, radius, the range of hydrogen-rich envelope mass, the range of explosion energy (defined as the energy of the thermal bomb minus the binding energy of the progenitor) and the range of $^{56}$Ni mass in log scale. The masses, luminosities and radii are in solar unit. The effective temperature $T_{\rm eff}$ is in the unit of K. The explosion energy is in the unit of foe (10$^{51}$ erg).}
\end{deluxetable*}

Variations in $M_{\rm ZAMS}$ should be understood as a means of generating models with different radii $R_{\rm prog}$ at RSG phases. Since $M_{\rm Henv}$ is treated as a free parameter, $R_{\rm prog}$ plays a more fundamental role in determining the light curve properties of SNe II than $M_{\rm ZAMS}$ (see \citealt{kasen09}; see also discussions in \S3.2). Given that our models maintain an almost constant $T_{\rm eff}\,\sim$\,3650\,K at the onset of the explosion, we use the luminosity of the RSG models, log\,$L_{\rm prog}$, as the proxy of $R_{\rm prog}$. In latter discussion, although we will use the term M10 to refer models with $M_{\rm ZAMS}$\,=\,10\,$M_{\rm \odot}$ for convenience, it actually represent models with log\,$L_{\rm prog}$\,=\,4.52 or $R_{\rm prog}\,\sim\,450\,R_{\rm \odot}$ (see the $M_{\rm ZAMS}$-log\,$L_{\rm prog}$-$R_{\rm prog}$ relation in Table~\ref{tab:model_grid}).

Using log\,$L_{\rm prog}$ rather than $M_{\rm ZAMS}$ as the primary variable has an important advantage: log\,$L_{\rm prog}$ can be directly measured from pre-SN images, and enters the light curve models via log\,$L_{\rm prog}\propto R_{\rm prog}^2$ at fixed, observation-constrained $T_{\rm eff}$. By contrast, $M_{\rm ZAMS}$ must be inferred indirectly, most commonly via the theoretically dependent $M_{\rm ZAMS}$-log\,$L_{\rm prog}$ relation (see \citealt{farrell20a} for difficulty in inferring $M_{\rm ZAMS}$ from log\,$L_{\rm prog}$), which is sensitive to the uncertain internal mixing prescriptions and other details of the stellar evolution models (see \citealt{farmer16,temaj24,fang25c}). By framing our discussions in terms of log\,$L_{\rm prog}$, we reduce these uncertainties, ensuring that our results are robust regardless of the assumed $M_{\rm ZAMS}$-log\,$L_{\rm prog}$ relations. The estimation of log\,$L_{\rm prog}$ and the associated uncertainties will be discussed in \S3.3.

\subsection{Pre-SN mass-loss}
The light curve grid is constructed from RSG models with the hydrogen-rich envelope artificially removed. This method allows direct control over $M_{\rm Henv}$ and enables the inference of its value by fitting the light curves of SNe II (\S3). Since our goal is to use the inferred $M_{\rm Henv}$ to constrain pre-SN mass-loss mechanisms, we do not impose any specific mass-loss model as a prior in the light curve modeling. However, once the $M_{\rm Henv}$ distribution and its connection with other physical parameters are established from observations, they can be used to compare with the predictions of various pre-SN mass-loss scenarios. To this end, we evolve additional model grids incorporating different stellar wind prescriptions and binary interactions that will be examined in \S4.

\subsubsection{Stellar wind mass-loss}
A grid of RSGs with $M_{\rm ZAMS}$ varying from 10 to 18\,$M_{\rm \odot}$ is evolved with the same set up as the models described in \S2.1, except that the hydrogen-rich envelope are not treated as a free parameter, but stripped by the stellar wind. When the star is hot ($T_{\rm eff}$\,\(>\)\,11000\,K), we use the \texttt{Vink} scheme (\citealt{vink01}; denoted as $\dot{M}_{\rm V01}$). At lower temperature ($T_{\rm eff}$\,\(<\)\,10000\,K), we adopt two cool wind mass-loss rates:
\begin{itemize}
    \item {\texttt{de Jager} scheme}. The \texttt{de Jager} mass-loss rate for cool RSGs is one of the most widely adopted mass-loss schemes. It is used in the \texttt{KEPLER} models and serves as the default prescription in the \texttt{MESA} code at low temperatures. This mass-loss rate (hereafter denoted as $\dot{M}_{\rm J88}$) is derived based on a sample of M-type stars, and can be empirically described as
    \begin{equation}
    \begin{aligned}
        &\quad{\rm log}(-\dot{M}_{\rm J88})\,=\,1.769\,{\rm log}\,L\\
        &\quad~-~1.676\,{\rm log}\,T_{\rm eff} - 8.158,
        \end{aligned}
    \end{equation}
where $\dot{M}_{\rm J88}$, $L$ and $T_{\rm eff}$ are in the units of $M_{\rm \odot}\,{\rm yr}^{-1}$, $L_{\rm \odot}$ and K respectively (\citealt{dejager88}). Note $L$ is the bolometric luminosity of the RSG, and should be distinguished from the terminal bolometric luminosity $L_{\rm prog}$ defined at the core carbon depletion.
    \item {\texttt{Yang} scheme}. In a recent work, \citet{yang23} derive the mass-loss rate for a large sample of M-type RSGs in the Small Magellanic Cloud (hereafter denoted as $\dot{M}_{\rm Y23}$)
    \begin{equation}
        \begin{aligned}
        &\quad{\rm log}(-\dot{M}_{\rm Y23})\,=\,0.45\,({\rm log}\,L)^3~\\
        &\quad-~5.26\,({\rm log}\,L)^2~+~20.93\,{\rm log}\,L~\\&\quad-~34.56,
        \end{aligned}
    \end{equation}
    where $\dot{M}_{\rm Y23}$ and $L$ are in the units of $M_{\rm \odot}\,{\rm yr}^{-1}$ and $L_{\rm \odot}$ respectively. To apply this relation in our models, we define its validity range based on $T_{\rm eff}$ following the method in \citet{zapartas24}: (1) For $T_{\rm eff}$\,\(>\)\,5000\,K, we adopt the \texttt{de Jager} scheme $\dot{M}_{\rm J88}$; (2) For $T_{\rm eff}$\,\(<\)\,4000\,K, we adopt the \texttt{Yang} scheme $\dot{M}_{\rm Y23}$; (3) In the transition range ($4000\,{\rm K} \leq T_{\rm eff} \leq 5000\,{\rm K}$), the linear interpolation (in terms of $T_{\rm eff}$) between $\dot{M}_{\rm J88}$ and $\dot{M}_{\rm Y23}$ is applied. Here we use a different transition range than \citet{zapartas24} to ensure that $\dot{M}_{\rm Y23}$ fully applies only when the star begins to climb the Hayashi track. During this phase, $T_{\rm eff}$ remains nearly constant, minimizing any intrinsic dependence of $\dot{M}_{\rm Y23}$ on $T_{\rm eff}$. Our own experiment shows that changing the transition range has a minimal effect on the results (see also \citealt{zapartas24}).
\end{itemize}
In the transition range ($10000\,{\rm K} \leq T_{\rm eff} \leq 11000\,{\rm K}$), the linear interpolation between $\dot{M}_{\rm J88}$ and $\dot{M}_{\rm V01}$ is applied.

\subsubsection{Binary model}
We evolve a grid of binary models using \texttt{MESA} with primary $M_{\rm ZAMS}$ varying from 10 to 15\,$M_{\rm \odot}$ (in 1\,$M_{\rm \odot}$ step) based on the test suite evolve\_both\_stars, modified following the setup of \citet{ercolino23} except that we use a larger $\alpha_{\rm MLT}$, which results in the same $T_{\rm eff}$ range as the RSG progenitors in this work. For numerical stability of systems with relatively low mass primaries, we adopt an exponential overshooting parameter $f_{\rm ov}$\,=\,0.018, which leads to a larger helium core mass (brighter log\,$L_{\rm prog}$) given the same $M_{\rm ZAMS}$, compared to the models in \S2.1. We use \texttt{de Jager} scheme to model the wind-driven mass-loss. The initial mass ratio of the primary to companion $q_{\rm i}$ ranges from 0.1 to 0.9 (in step 0.1), and the initial separation log\,$R_{\rm sep}/R_{\rm \odot}$ ranges from 3.0 to 3.5\,dex (in step 0.02 dex). Both the primary and companion stars have solar metallicity, and the initial rotations are relaxed to the orbital period at the beginning of the evolution.

In this setup, the companion is not treated as a point mass, but the two stars are evolved simultaneously. The relatively wide separation ensures that the two stars remain detached during main-sequence phase of the primary star. Once the primary star enters the RSG phase and expands to fill its Roche lobe, we use the \texttt{Kolb} scheme to model the binary mass transfer (\citealt{kolb}). The accretion of the mass flow onto the companion, and the resultant angular momentum transport, are modeled following \citet{deMink13}. The mass transfer efficiency, defined by the ratio of the mass accreted onto the companion and the mass of the accretion flow, is determined by the spin of the accretor (\citealt{deMink13,ercolino23,fragos23}). After a few fractions of a solar mass is accreted, the companion is spun-up to critical rotation by the gained angular momentum \citep{packet81}. Further accretion is restricted and the mass flow from the binary mass transfer is ejected from the system by rotation-enhanced stellar wind (\citealt{heger98,langer98}; but see \citealt{popham91} that accretion may continue through star-disk interaction).

The calculation is carried out until core carbon depletion of the primary star, or when unstable mass transfer occurs. We employ two criteria as unstable mass transfer: (1) that the mass-loss rate $\dot{M}$ exceeds 0.1\,$M_{\rm \odot}\,{\rm y}r^{-1}$. This is a simplified criterion, though \citet{fragos23} show that $\dot{M}$ in such systems can eventually reach arbitrarily high rates; (2) that the radius of the primary star fills the outer Lagrangian point located at $\sim$\,1.3\,$R_{\rm L}$ \citep{pavlovskii15,marchant16,misra20,marchant21}. Here $R_{\rm L}$ is the Roche lobe radius. Mass-loss through this point carries a large amount of angular momentum, and the two stars start to spiral in, which may lead to common envelope evolution (\citealt{henneco24,xing24}). If one of these criteria is met, the calculation is stopped. In this work, the analysis is restricted to stable binary mass transfer. The MESA inlists used to generate the binary models are available on Zenodo under an open-source Creative Commons Attribution 4.0 International license: doi:10.5281/zenodo.15744709.

\bigskip 
In \S3, We will first use the model grid constructed in \S2.1 to derive the physical parameters from SNe II light curve modeling. In \S4, the derived parameters and their mutual relations are compared with the predictions of the stellar wind models (\S2.2.1) and binary models (\S2.2.2) to constrain the pre-SN mass-loss mechanism.

\section{Inferring physical parameters from light curve modeling}
The goal of this work is to derive the physical properties of SNe II and their RSG progenitors with additional constraints from nebular spectroscopy. To achieve this, we select SNe II from the nebular spectroscopy sample of \citet{fang25c} that also have multi-band photometry covering both the plateau and nebular phases. Based on these selection criteria, our final sample consists of 32 SNe II, listed in Table~\ref{tab:sample_appendix}. We use the \texttt{Python} package \texttt{emcee} (\citealt{emcee}) to infer the optimal parameters for the observed light curves using the model grid constructed in \S2.1. In \S3.1, we introduce the light curve interpolation method that is necessary for this approach. As we will show in \S3.2, the physical parameters inferred from light curve modeling exhibit a degeneracy with log\,$L_{\rm prog}$ (or $M_{\rm ZAMS}$), similar to the degeneracy with $R_{\rm prog}$ reported by \citet{goldberg20}. In \S3.3, we demonstrate how nebular spectroscopy can be employed to resolve this degeneracy.

\subsection{Model interpolation}
\begin{figure}
\epsscale{0.88}
\plotone{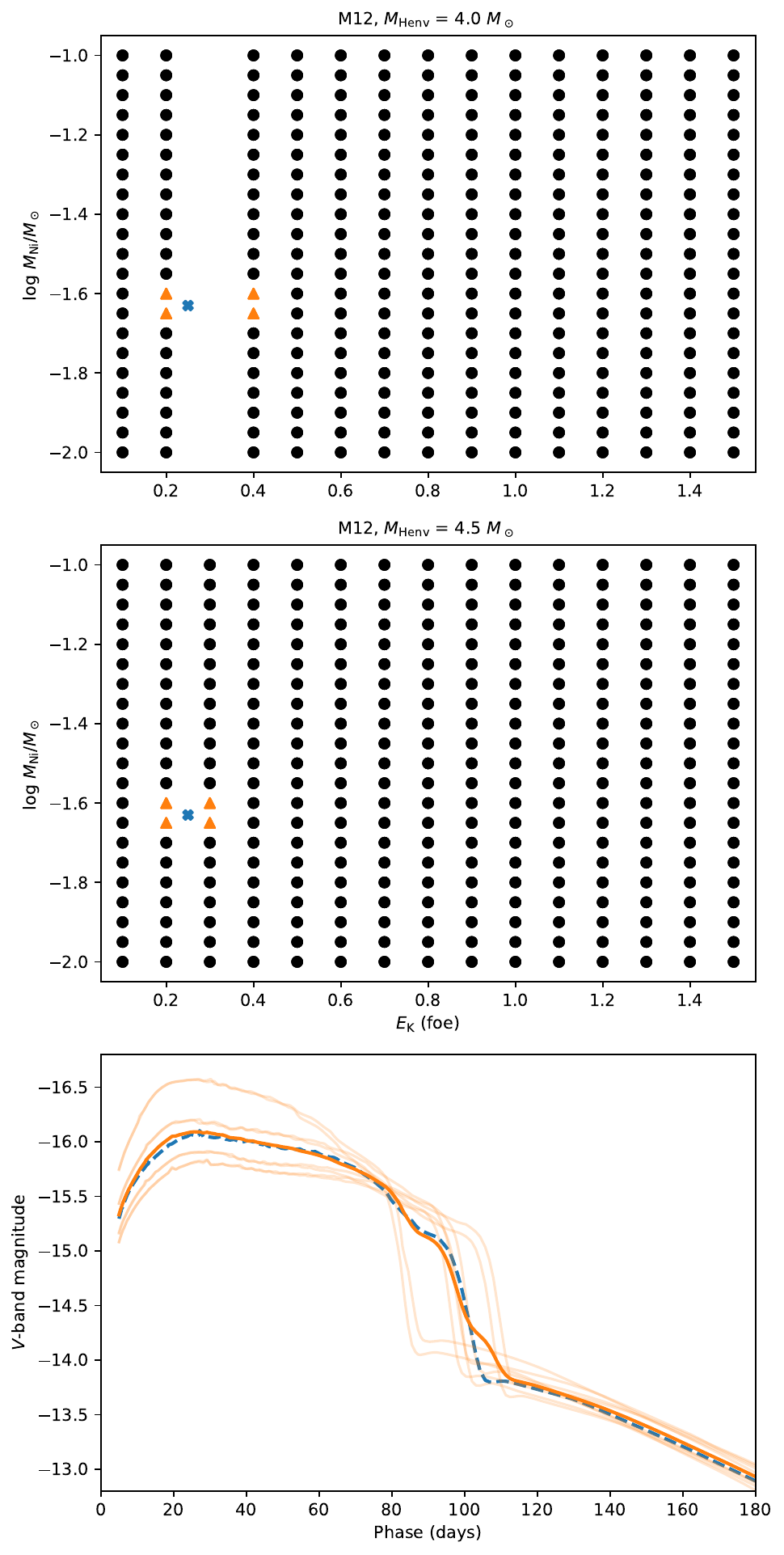}
\centering
\caption{The interpolation to $\theta$\,=\,\{4.3, 0.25, -1.63\}, and log\,$L_{\rm prog}$\,=\,4.72 (M12 model). The upper and middle panels illustrate the pre-computed grid points of $E_{\rm K}$ and log\,$M_{\rm Ni}$ of M12 models with $M_{\rm Henv}$\,=\,4.0 and 4.5\,$M_{\rm \odot}$, respectively. The blue cross is the position of $\theta$\,=\,\{4.3, 0.25, -1.63\} on the $E_{\rm K}$-log\,$M_{\rm Ni}$ plane, and the orange triangles are the grid points that contribute to the interpolation ($\theta_{\rm i}\,\in\,\theta_{\rm close}$). Lower panel: The interpolated $V$-band light curve. The orange solid line is the result of the interpolation of $V$-band magnitudes, and the transparent orange lines are the $V$-band light curves of the models in $\theta_{\rm close}$. The blue dashed line is the model calculated by \texttt{STELLA} with the same $\theta$ and log\,$L_{\rm prog}$\,=\,4.72 (M12 model).}
\label{fig:interpolation}
\end{figure}

Before performing the parameter inference with \texttt{emcee}, we need to establish a method to predict the light curves for models with arbitrary parameter combinations, which may not be covered by the pre-computed grid points. To achieve this, we adopt an interpolation method similar to that used in \citet{F18}: For a fixed log\,$L_{\rm prog}$ ($M_{\rm ZAMS}$), the intrinsic physical parameters are defined to be $\theta$\,=\,\{$M_{\rm Henv}$, $E_{\rm K}$, log\,$M_{\rm Ni}$\}. In later section, we will explain why log\,$L_{\rm prog}$ is treated separately from $\theta$. The interpolated magnitude at phase $t$, denoted as $m$($t,\theta$), is then calculated as
\begin{equation}
     m(t,\theta)\,=\,\sum_{\theta_{i}\,\in\,\theta_{\rm close}} m(t,\theta_{i})\,\times\,\hat{w}(\theta,\theta_{i}).
\end{equation}
Here, $m(t,\theta_{\rm i})$ is the magnitude of the model with fixed log\,$L_{\rm prog}$ and parameter $\theta_{\rm i}$ in the grid, and $\theta_{\rm close}$ is the set of $\theta_{\rm i}$ that surround $\theta$, as graphically illustrated in Figure~\ref{fig:interpolation}. The normalized weighting function $\hat{w}(\theta,\theta_{\rm i})$ is
\begin{equation}
     \hat{w}(\theta,\theta_{i})\,=\,\frac{w(\theta,\theta_{i})}{\sum_{\theta_{j}\,\in\,\theta_{\rm close}} w(\theta,\theta_{j})}.
\end{equation}
In this work, we use a different weighting function $w(\theta,\theta_{j})$ than \citet{F18}, defined as:
\begin{equation}
     w(\theta,\theta_{i})\,=\, \left[ \sqrt{\sum_{k=1,2,3}(\frac{\theta^{k} - \theta^{k}_{i}}{\Delta\theta^{k}})^2} + 10^{-6}\right]^{-1}.
\end{equation}
Here $\theta^{k}$ is the $k$-th parameter of $\theta$ (for example, $\theta^{1}$ is $M_{\rm Henv}$), and $\Delta\theta^{k}$ is the maximum difference of the $k$-th parameter for all elements in $\theta_{\rm close}$. Therefore $w(\theta,\theta_{j})$ is the inverse of the $relative$ distances of the parameters in $\theta$ and $\theta_{i}$. The term 10$^{-6}$ is attached to avoid divergence.

To assess the quality of the interpolation, we compute multi-band light curves for $\theta$\,=\,\{4.3, 0.25, -1.63\} with log\,$L_{\rm prog}$\,=\,4.72 (M12 models), a parameter set that is not among the pre-computed grid points, and compare them to direct radiative hydrodynamics simulation with \texttt{STELLA} for the same parameters. In the upper and lower panels of Figure~\ref{fig:interpolation}, the pre-computed grid models that contribute to the interpolation, i.e., $\theta\,\in\,\theta_{\rm close}$, are labeled by the orange triangles. In this case, the parameter intervals are $\Delta \theta^{1}$\,=\,0.5\,$M_{\rm \odot}$, $\Delta \theta^{2}$\,=\,0.2\,foe and $\Delta \theta^{3}$\,=\,0.05\,dex. The interpolated $V$-band light curve is shown as the orange solid line in the lower panel, with the light curves of the surrounding grid models ($\theta\,\in\,\theta_{\rm close}$) overlaid as transparent lines. The light curve from radiative hydrodynamics simulation is plotted as the blue dashed line. The interpolation performs well on the plateau phase and the radiative tail, with deviations between these two methods generally within 0.05 mag. However, the accuracy becomes worse during the transition phase, where the light curve shows a steep gradient that is difficult to be captured by the employed interpolation method. Nevertheless, since the transition phase is short and the observed data points are typically sparsely sampled in this regime, the reduced interpolation accuracy within it does not significantly affect the overall results of this work.

\subsection{Fitting the light curves}
With the interpolation method, we can rapidly compute the light curves with arbitrary combinations of $\theta$, allowing for using the \texttt{emcee} routine to infer the optimized $\theta$ for the observed light curve with 
a fixed log\,$L_{\rm prog}$ ($M_{\rm ZAMS}$). For each log\,$L_{\rm prog}$ value, besides the 3 physical parameters \{$M_{\rm Henv}$, $E_{\rm K}$, log\,$M_{\rm Ni}$\}, we include another parameter $\sigma_{\mu}$ to characterize the uncertainty in distance modulus. The priors of the physical parameters follow the uniform distributions within their ranges shown in Table~\ref{tab:model_grid}. For $\sigma_{\mu}$, its prior follows a Gaussian distribution with standard deviation listed in Table~\ref{tab:sample_appendix}. With this setup, we derive the posterior distributions of the parameters by modeling the multi-band light curves with \texttt{emcee} when log\,$L_{\rm prog}$ is fixed. Since CSM is not included in the models, photometry for the first 30 days is removed from the fitting procedure. Figure~\ref{fig:interpolation} shows the projected corner plots for SN 2013by with log\,$L_{\rm prog}$\,=\,4.96 (M15 models). 

\begin{figure}
\epsscale{1.2}
\plotone{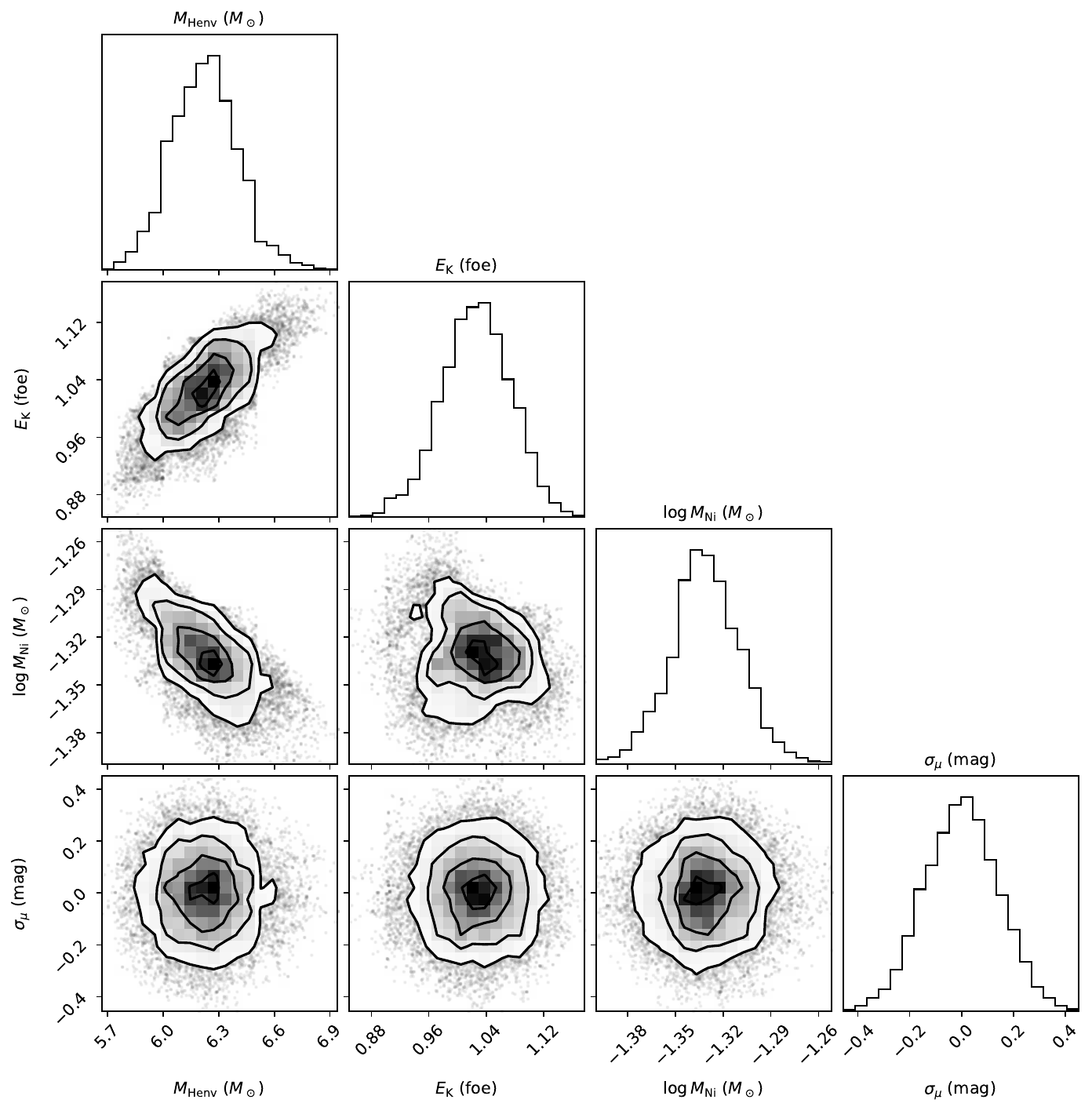}
\centering
\caption{The corner plot of the posterior distributions of the parameters \{$M_{\rm Henv}$, $E_{\rm K}$, log\,$M_{\rm Ni}$, $\sigma_{\mu}$\} obtained using the \texttt{emcee} routine for SN 2013by, with fixed log\,$L_{\rm prog}$\,=\,4.96 (M15 models).}
\label{fig:2014G_corner}
\end{figure}

In principle, we can apply the same method to simultaneously optimize \{log\,$L_{\rm prog}$, $M_{\rm Henv}$, $E_{\rm K}$, log\,$M_{\rm Ni}$\} (see \citealt{hiramatsu21} for example). However, as discussed in \citet{dessart19} and \citet{fang25a}, once $M_{\rm Henv}$ is allowed to vary freely, the effect of $M_{\rm ZAMS}$, therefore log\,$L_{\rm prog}$, on light curve becomes relatively weak compared with $M_{\rm Henv}$, $E_{\rm K}$ and log\,$M_{\rm Ni}$. Including log\,$L_{\rm prog}$ as a free parameter may introduce unnecessary complexity into the model. 

Indeed, despite the diversity, the most important properties of the light curves are: (1) the plateau magnitude, (2) the plateau duration, and (3) the magnitude of the radioactive tail (\citealt{khatami19}). These quantities are mainly determined by 4 parameters: (1) the progenitor radius $R_{\rm prog}$, (2) the hydrogen-rich envelope mass $M_{\rm Henv}$, (3) the explosion energy $E_{\rm K}$ and (4) the $^{56}$Ni mass in the ejecta log\,$M_{\rm Ni}$ (\citealt{popov93,kasen09,goldberg19,fang25a}). In a simplified view, for a fixed log\,$L_{\rm prog}$ (therefore fixed progenitor radius $R_{\rm prog}$ for models in this work), $E_{\rm K}$ and log\,$M_{\rm Ni}$ govern the magnitudes of the plateau and radioactive tail respectively (see Figure~10 of \citealt{fang25a} that the plateau magnitude of $V$-band light curve is almost constant when $E_{\rm K}$ is fixed at 1\,foe). Once $E_{\rm K}$ and log\,$M_{\rm Ni}$ are determined, $M_{\rm Henv}$ can be inferred from the plateau duration, reducing the problem to solving 3 unknowns with 3 constraints—a closed system. Introducing log\,$L_{\rm prog}$ ($R_{\rm prog}$) as a free parameter increases the number of unknowns without adding new constraints, making the system under-determined. For practical purposes, excluding log\,$L_{\rm prog}$ from $\theta$ helps maintain a well-constrained model. In \S5.2, we will discuss the results when log\,$L_{\rm prog}$ is also included in the \texttt{emcee} routine.

\begin{figure*}
\epsscale{1.12}
\plotone{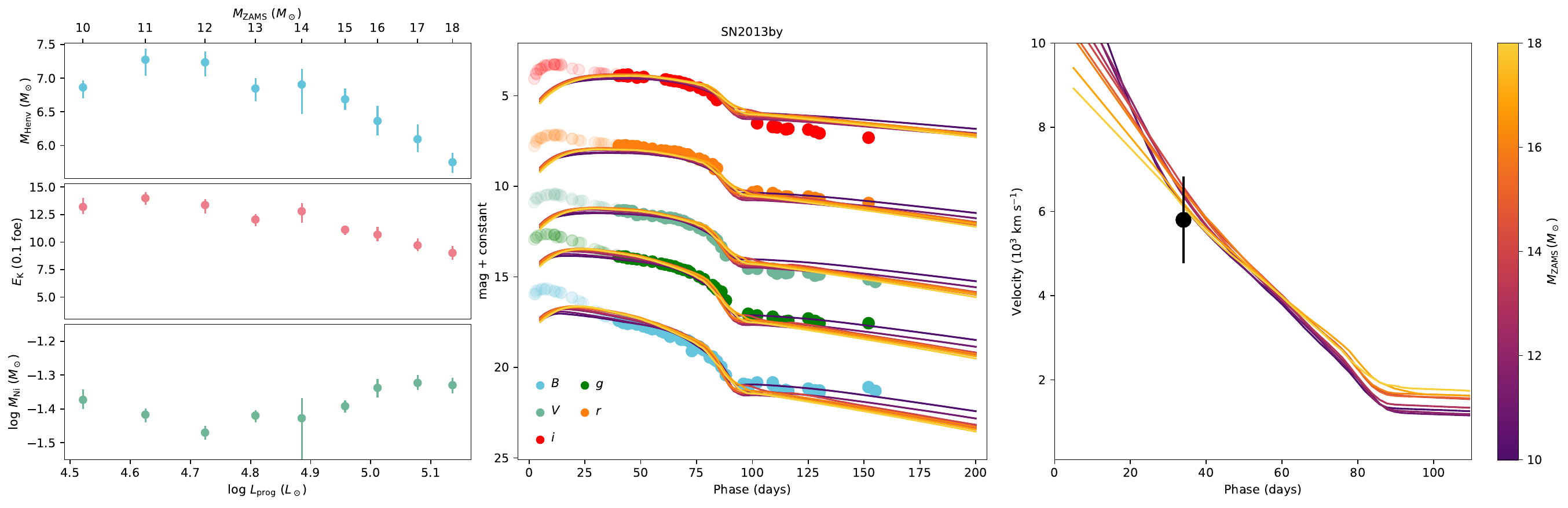}
\centering
\caption{The results of the \texttt{emcee} inference for the multi-band light curves of SN 2013by. Left panels: The optimized $M_{\rm Henv}$, $E_{\rm K}$ and log\,$M_{\rm Ni}$ as functions of log\,$L_{\rm prog}$ ($M_{\rm ZAMS}$). The error bars represent the 68\% CIs of the posterior distributions. Middle panel: The optimized $V$-band light curves for SN 2013by, with different fixed log\,$L_{\rm prog}$ ($M_{\rm ZAMS}$) indicated by the color-bar. Despite the models have a broad range of log\,$L_{\rm prog}$, there is no visual difference in their optimized light curves; Right panel: velocities of the optimized models, measured at the mass coordinate where the Sobolev optical depth $\tau_{\rm Sob}$ of Fe II\,$\lambda$5169 equals unity. The black scatter point is the Doppler velocity of the Fe II $\lambda$5169, measured for SN 2013by at $t$\,=\,34\,days after the explosion.}
\label{fig:2014G_example}
\end{figure*}

In some studies, photospheric velocity $v_{\rm ph}$, measured from Doppler shifts of the absorption features in early phase spectroscopy, is used as an additional constraint (see, e.g., \citealt{martinez19,martinez20,martinez22}). However, photospheric velocity is not truly independent. Observationally, a correlation exists between plateau luminosity and the Doppler velocity of Fe II\,$\lambda$5169, both measured at 50 days post-explosion (\citealt{hamuy03}). This correlation is explained by \citet{kasen09}, where they show that $v_{\rm ph}$ merely traces the luminosity of the plateau at the same phase and does not provide an independent constraint. This degeneracy was further confirmed by a survey of a large sample of \texttt{MESA}\,+\,\texttt{STELLA} models in \citet{goldberg19}, and they concluded that $v_{\rm ph}$ should not be used to infer explosion properties (see also \citealt{fang25a}).

In the left panels of Figure~\ref{fig:2014G_example}, we show the dependence of the optimized \{$M_{\rm Henv}$, $E_{\rm K}$, log\,$M_{\rm Ni}$\} on log\,$L_{\rm prog}$, where a clear trend emerges: as log\,$L_{\rm prog}$ increases, the optimized $M_{\rm Henv}$ and $E_{\rm K}$ tend to decrease, while log\,$M_{\rm Ni}$ shows a slight increase. This behavior follows from the scaling relations of SNe II light curve properties (\citealt{kasen09,goldberg19,fang25a}): increasing log\,$L_{\rm prog}$ effectively corresponds to increasing the radius of the progenitor RSG (see Table~\ref{tab:model_grid}), which, if all other parameters remain fixed, leads to a brighter plateau. For SNe II from luminous RSGs, achieving the same plateau magnitude requires a lower $E_{\rm K}$, resulting in longer plateau. Consequently, a smaller $M_{\rm Henv}$ is needed. 

The behavior of log\,$M_{\rm Ni}$ is more complex: although both the optimized $E_{\rm K}$ and $M_{\rm Henv}$ decrease with increasing log\,$L_{\rm prog}$, the $\gamma$-ray diffusion timescale, which roughly scales as
\begin{equation}
    T_{\rm 0}\,\propto\,\frac{M_{\rm Henv}}{E_{\rm K}^{1/2}},
\end{equation}
also decreases (\citealt{clocchiatti97,wheeler15,haynie23}). The $\gamma$-ray photons from the radioactive decay chain are therefore less trapped. To achieve the same magnitude on the radioactive tail, larger log\,$M_{\rm Ni}$ is needed. 

In the middle panel of Figure~\ref{fig:2014G_example}, we assess the quality of the fits by computing interpolated light curves using the parameter set $\theta$ that minimizes the likelihood function for each log\,$L_{\rm prog}$. The results show that all values of log\,$L_{\rm prog}$ produce similarly good fit, with no significant differences in fit quality. Once log\,$L_{\rm prog}$ is fixed, the light curve properties are well determined by \{$M_{\rm Henv}$, $E_{\rm K}$, log\,$M_{\rm Ni}$\}. 

In the right panel of Figure~\ref{fig:2014G_example}, we compare the velocities of the optimized models, measured at the mass coordinate where the Sobolev optical depth $\tau_{\rm Sob}$ of Fe II $\lambda$5169 equals unity, with the observed Doppler velocity of Fe II $\lambda$5169 in SN 2013by measure for the spectroscopy taken on 2013 May 26 ($t = 34$ days; \citealt{valenti_13by}). Despite log\,$L_{\rm prog}$ varying from 4.52 to 5.13 ($M_{\rm ZAMS}$ from 10 to 18\,$M_{\odot}$), the optimized parameters \{$M_{\rm Henv}$, $E_{\rm K}$, log\,$M_{\rm Ni}$\} for different fixed log\,$L_{\rm prog}$, inferred solely from light curve modeling, yield photospheric velocities that closely match the observations (see also \citealt{fang25b}). This further confirms our earlier discussion that $v_{\rm ph}$ does not provide an independent constraint for breaking the degeneracies between \{$M_{\rm Henv}$, $E_{\rm K}$, log\,$M_{\rm Ni}$\} and log\,$L_{\rm prog}$.

\subsection{Additional constraints from nebular spectroscopy}
\begin{figure*}
\epsscale{1.}
\plotone{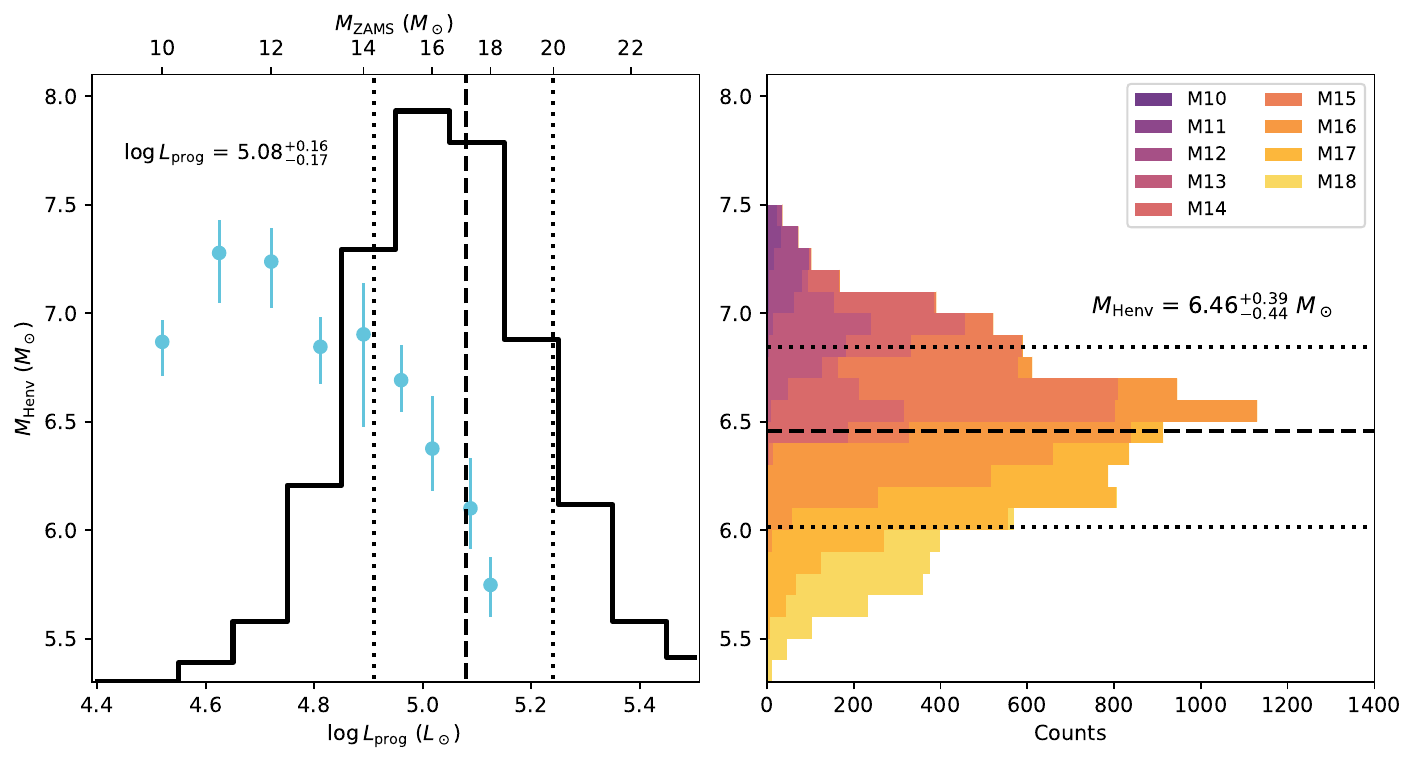}
\centering
\caption{The posterior $M_{\rm Henv}$ of SN 2013by, weighted by log\,$L_{\rm prog}$ of the progenitor RSG. Left panel: The light blue scatter points represent the median values and 68\% CIs of the posterior $M_{\rm Henv}$ from light curve modeling as function of log\,$L_{\rm prog}$ ($M_{\rm ZAMS}$). The black histogram represents the log\,$L_{\rm prog}$ probability density function inferred from nebular spectroscopy, following the method outlined in \citet{fang25c}. Right panel: The $M_{\rm Henv}$ distribution weighted by log\,$L_{\rm prog}$ (see the main text for the process). The contributions from $M_{\rm Henv}$ distributions, with different fixed log\,$L_{\rm prog}$, are color-coded.}
\label{fig:weight}
\end{figure*}

Although the light curve of SNe II contains limited information of log\,$L_{\rm prog}$, the [O I] emission in nebular spectroscopy offers additional insights. In principle, comparing observed nebular spectra with models (\citealt{jerkstrand12, jerkstrand14}) allows for direct inference of $M_{\rm ZAMS}$, which can be converted to log\,$L_{\rm prog}$ through the $M_{\rm ZAMS}$-log\,$L_{\rm prog}$ relations of the stellar evolution models. By combining this with the dependence of \{$M_{\rm Henv}$, $E_{\rm K}$, log\,$M_{\rm Ni}$\} on log\,$L_{\rm prog}$ established in the previous section, we can derive final estimates for these parameters. However, caution is required. As discussed earlier, log\,$L_{\rm prog}$ affects the properties of the light curve through its relation with the radius of the RSG progenitor $R_{\rm prog}$ given that the models are restricted to a limited range of $T_{\rm eff}$. In contrast, the oxygen mass $M_{\rm O}$ is irrelevant to this aspect. Since the [O I] line strength is mainly determined by $M_{\rm O}$, inferring $M_{\rm ZAMS}$ and hence log\,$L_{\rm prog}$ from nebular spectroscopy relies on the $M_{\rm O}$-log\,$L_{\rm prog}$ relation, which is affected by many factors such as the mixing of the material during stellar evolution (\citealt{temaj24}). In \citet{fang25c}, a comparison of different stellar evolution models and mixing assumptions reveals significant scatter in this relation (see the lower panel of Figure 8 therein). As a result, $M_{\rm ZAMS}$ measured from nebular spectroscopy (denoted as $M_{\rm ZAMS,neb}$) does not directly corresponds to log\,$L_{\rm prog}$, the key parameter in light curve modeling.

In \citet{fang25c}, an approach to infer log\,$L_{\rm prog}$ from nebular phase spectroscopy is developed. The readers may refer to that paper for full details, and we briefly describe the work flow as follows:
\begin{itemize}
    \item For each SNe II, we first normalize the nebular spectroscopy by the integrated flux within the wavelength range 5000 to 8500\,{\rm{\AA}}, then measure the [O I] fractional flux $f_{\rm [O\,I]}$ and its regulated form $f_{\rm [O\,I],reg}$ where H$\alpha$ is subtracted from the integrated flux to account for the uncertainties in $M_{\rm Henv}$ and the contributions from shock-CSM interaction. The corresponding $M_{\rm ZAMS,neb}$ is determined by comparing observed $f_{\rm [O\,I]}$ ($f_{\rm [O\,I],reg}$) with the models of \citet{jerkstrand12,jerkstrand14} at the same phase. If an object have $f_{\rm [O\,I]}$ ($f_{\rm [O\,I],reg}$) below the M12 models, the only information here is that it has low progenitor core mass. The corresponding $M_{\rm ZAMS,neb}$ is not a single value with Guassian uncertainties, but is assumed to be a flat distribution within 10 to 12\,$M_{\rm \odot}$;
    \item Within our sample, 13 SNe II have pre-SN images and measured log\,$L_{\rm prog}$. For these objects (golden sample), a strong correlation between $M_{\rm ZAMS,neb}$-log\,$L_{\rm prog}$ is discerned (SN 2013ej is an outlier; see discussions therein), which is used to estimate log\,$L_{\rm prog}$ from $M_{\rm ZAMS,neb}$ for other objects without pre-SN images;
    \item We generate 10,000 trials in which $M_{\rm ZAMS,neb}$ and (where available) log\,$L_{\rm prog}$ are randomly perturbed by their uncertainties. Each trial yields a new linear regression between $M_{\rm ZAMS,neb}$-log\,$L_{\rm prog}$ from the golden sample, which is applied to all other SNe II.
    \item After 10,000 trials every SN now has a distribution of log\,$L_{\rm prog}$ values. The median value and the 68\% confidence interval (CI; defined by the 16th and 84th percentiles) of this log\,$L_{\rm prog}$ distribution for each object is listed in Table~\ref{tab:sample_appendix}, and the probability density function is denoted as $P$(log\,$L_{\rm prog}$).
\end{itemize}

Although the values of $M_{\rm ZAMS,neb}$ depend on the chosen nebular model grid, the resulting log\,$L_{\rm prog}$, which will be employed to break the degeneracy in light curve modeling, is not sensitive to the spectral or stellar evolution models, as it is ultimately anchored to log\,$L_{\rm prog}$ measured from the observed pre-SN images (see \S4.3 in \citealt{fang25c}). The key assumption of this approach is that all SNe II follow a universal $M_{\rm ZAMS,neb}$-log\,$L_{\rm prog}$ relation calibrated by the golden sample, which will require a larger sample of SNe II that both nebular spectroscopy and pre-SN imaging are available to test its validity. Another potential source of systematic uncertainty arises from the fact that our light curve models are based on RSG progenitors computed with MESA, while the nebular spectral models are based on KEPLER models. However, as shown in Figure 8 of \citet{fang25c}, the $M_{\rm He\,core}$-log\,$L_{\rm prog}$ relations predicted by these two codes are in good agreement within the core mass range considered in this work. Therefore, despite the use of different stellar evolution codes, the weighting process introduced below is not significantly affected.

Combined with the dependence of \{$M_{\rm Henv}$, $E_{\rm K}$, log\,$M_{\rm Ni}$\} on log\,$L_{\rm prog}$, the log\,$L_{\rm prog}$ distribution constructed above is used to establish the marginal posterior distributions of these parameters as:
\[
    P(x)\,=\,\int\int P(x|{\rm log}\,L_{\rm prog})\,P({\rm log}\,L_{\rm prog})d\,{\rm log}\,L_{\rm prog}\,dx,
\]
where $x$ represents one of the parameter in \{$M_{\rm Henv}$, $E_{\rm K}$, log\,$M_{\rm Ni}$\}, and $P(x|{\rm log}\,L_{\rm prog})$ is the posterior probability density function of $x$ at fixed log\,$L_{\rm prog}$ as constructed in \S3.2. The inferred values and associated uncertainties for these parameters are taken to be the median and 68\% CI of the resulting distribution $P(x)$. In practice, the procedure is carried out as described below:
\begin{itemize}
        \item For each individual SNe, a random value \( l \) is drawn from its \(\log L_{\rm prog}\) distribution established in \citet{fang25c}.
        \item Since $l$ may not match exactly with the pre-computed models in Table~\ref{tab:model_grid}, the corresponding $M_{\rm Henv}$ is estimated via interpolation. We identify the group number \( i \) such that  
        \[
        \log L_{{\rm prog},i} \,< \,l \,< \,\log L_{{\rm prog},i+1}.
        \]  
        For example, if $l\,$=\,4.85, then we find $i$\,=\,3, and log$\,L_{{\rm prog},i}$ and log$\,L_{{\rm prog},i+1}$ corresponds to M13 and M14 models respectively (see Table~\ref{tab:model_grid}).
        \item A value \( m_i \) is randomly drawn from the posterior distribution of $M_{\rm Henv}$ at fixed log$\,L_{{\rm prog},i}$, as established in \S3.2.
        \item Similarly, we obtain \( m_{i+1} \) from the posterior distribution of $M_{\rm Henv}$ at fixed log$\,L_{{\rm prog},i+1}$.
        \item The hydrogen-rich envelope mass $m$ corresponding to \( l \) is then estimated via linear interpolation
        \[
        m\,=\,\alpha\,\times\,m_{i+1}\,+\,(1-\alpha)\times\,m_{i},
        \]
        where
        \[
        \alpha\,=\,\frac{l - \log\,L_{{\rm prog},i}}{\log\,L_{{\rm prog},i+1} - \log\,L_{{\rm prog},i}}.
        \]
        In cases that $l$\,\(<\)\,4.52\,dex (M10 models), $m$ is estimated from the extrapolation of the M10 and M11 models.
    \item The previous steps are repeated 10,000 times to generate the final posterior distribution of \( M_{\rm Henv} \).
\end{itemize}
The $M_{\rm Henv}$ distribution established in this way, referred to as the log\,$L_{\rm prog}$-weighted $M_{\rm Henv}$ distribution throughout this work, naturally incorporates all sources of uncertainty. The Monte Carlo sampling method not only preserves the posterior distributions of $M_{\rm Henv}$ derived from light curve modeling when log\,$L_{\rm prog}$ is fixed, but also retains the shape of the log\,$L_{\rm prog}$ distribution established from nebular spectroscopy. As a result, the inferred parameters remain consistent with constraints from pre-SN images of the RSG progenitor, light curve modeling, and nebular spectroscopy.

The log\,$L_{\rm prog}$-weighted $M_{\rm Henv}$ distribution of SN 2013by, constructed from the above procedure, is shown in Figure~\ref{fig:weight} as example. The left panel illustrates the dependence of $M_{\rm Henv}$ on the fixed log\,$L_{\rm prog}$, and the distribution of log\,$L_{\rm prog}$ derived in \citet{fang25c} for this object is shown as the black histogram. The log\,$L_{\rm prog}$ weighted $M_{\rm Henv}$ distribution is shown in the right panel, with the contributions for different fixed log\,$L_{\rm prog}$ ($M_{\rm ZAMS}$) values distinguished by colors. 

Since the primary focus of this work is the pre-SN mass-loss mechanism in SNe II, our discussion is limited to $M_{\rm Henv}$. However, the distributions of $E_{\rm K}$ and log\,$M_{\rm Ni}$, along with their potential correlations with log\,$L_{\rm prog}$, are crucial for understanding the core-collapse mechanism, which will be explored in a forthcoming work (Fang et al., in preparation).

\section{Results}
In this section, we explore the pre-SN mass-loss mechanisms of SNe II by examining the relationship between $M_{\rm Henv}$ and log\,$L_{\rm prog}$. Since the sample of 32 SNe II in this work is biased with unknown selection function, we focus on the overall scatter rather than a detailed population analysis. 

Although the RSG mass-loss rate driven by stellar winds carries significant uncertainties, it primarily scales with log\,$L_{\rm prog}$, with more luminous RSGs tending to lose more mass through this channel (see \citealt{yang23} for example). In contrast, binary interactions can introduce substantial deviations from this trend. While there is evidence that more massive stars are more likely to be born in binary systems \citep{moe17, moe19}, the extent of mass stripping is primarily determined by orbital parameters, such as the binary separation and the mass ratio between the primary and companion stars. Consequently, binary evolution can lead to a wide range of hydrogen-rich envelope masses for progenitors with similar luminosities. 

\begin{figure}
\epsscale{1.}
\plotone{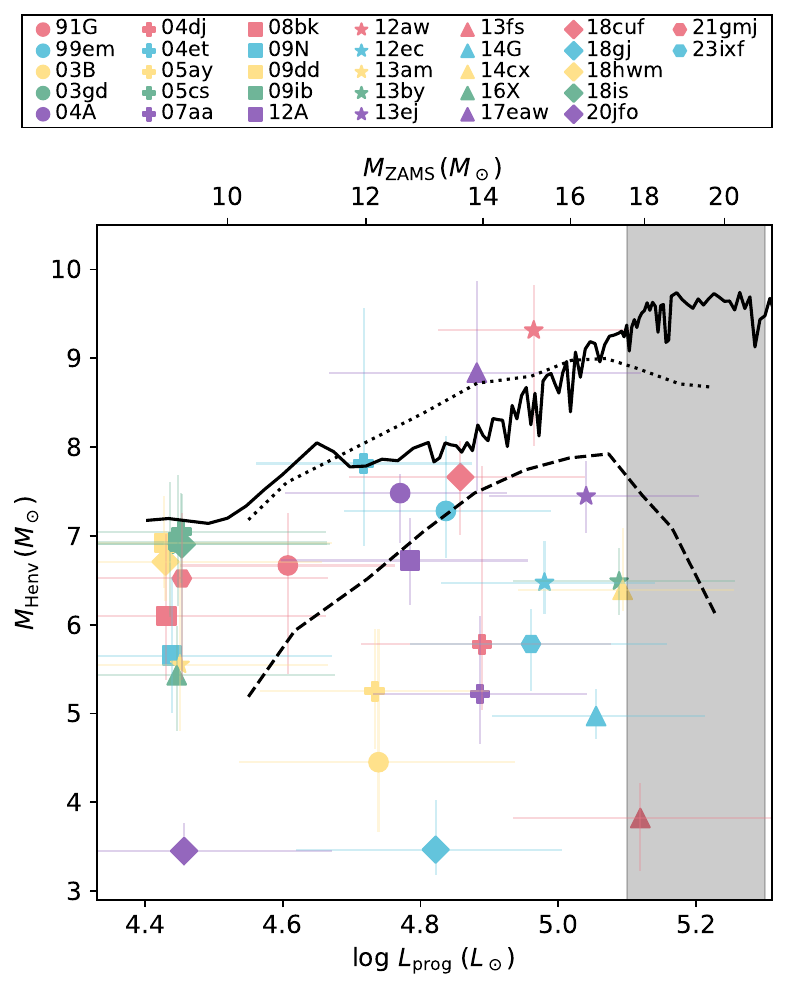}
\centering
\caption{The comparison between the progenitor luminosity log\,$L_{\rm prog}$, inferred from nebular spectroscopy, and the $M_{\rm Henv}$ inferred from light curve modeling. The error bars are the 68\% CIs of the posterior distributions, and different individual SNe are represented by different markers and colors. The black line are the log\,$L_{\rm prog}$-$M_{\rm Henv}$ relation predicted by different stellar evolution models: \texttt{KEPLER} models (solid), \texttt{MESA} models evolved with \texttt{de Jager} scheme (dotted) and \texttt{MESA} models evolved with \texttt{Yang} scheme (dashed). The shaded region represents log\,$L_{\rm prog}$\,=\,5.20$^{+0.10}_{-0.10}$, the upper luminosity cutoff of SNe II progenitors reported by several works (\citealt{smartt09,smartt15,DB20,fang25c}).}
\label{fig:main}
\end{figure}

In Figure~\ref{fig:main}, we compare log\,$L_{\rm prog}$, inferred from nebular phase spectroscopy (\citealt{fang25c}), with the log\,$L_{\rm prog}$-weighted $M_{\rm Henv}$ (\S3.3), derived from light curve modeling. The error bars indicate the 68\% CI of the distributions. Two key trends emerge from this comparison: (1) In nearly all cases, the inferred $M_{\rm Henv}$ values fall below the track predicted by the \texttt{KEPLER} models (black solid line), suggesting that most SNe II have experienced a greater mass loss than expected from single-star wind evolution alone; (2) low $M_{\rm Henv}$ values are observed across all luminosity ranges, indicating that the partial removal of the hydrogen-rich envelope is not solely determined by log\,$L_{\rm prog}$.

This empirical log\,$L_{\rm prog}$-$M_{\rm Henv}$ relation can be employed to constrain pre-SN mass-loss of the RSG progenitors. In this section, we investigate whether alternative mass-loss mechanisms can explain this observed scatter shown in Figure~\ref{fig:main}. Specifically, we explore the effects of different stellar wind prescriptions (\S4.1) and binary interactions (\S4.2). Additionally, we examine whether assumptions regarding the compactness of the RSG progenitor's envelope contribute to the discrepancy between models and observation (\S4.3). Since \texttt{KEPLER} progenitor models are widely used as initial conditions for modeling SN II light curves \citep{morozova18,moriya23,vartanyan25}, nebular spectroscopy \citep{jerkstrand12,jerkstrand14}, and core-collapse processes \citep{sukhbold18,ertl20,patton20,sukhbold20,burrows21,burrows24a,burrows24b}, they serve as the standard single-star models evolved with wind-driven mass loss in our discussions of \S4.2 and \S4.3.

\subsection{Stellar wind prescriptions}
In \S2.2.1, we have evolved a grid of RSG progenitors using the wind mass-loss of \texttt{de Jager} scheme (\citealt{dejager88}) and \texttt{Yang} scheme (\citealt{yang23}). The log\,$L_{\rm prog}$-$M_{\rm Henv}$ relation predicted by these cool stellar wind prescriptions are shown as the black dashed and dotted lines in Figure~\ref{fig:main}. In the range of log\,$L_{\rm prog}\,\leq$\,5.1, the models evolved with \texttt{de Jager} scheme follow a similar log\,$L_{\rm prog}$-$M_{\rm Henv}$ relation as the \texttt{KEPLER} models, however, the \texttt{Yang} scheme predicts a systematically lower $M_{\rm Henv}$. The difference between the \texttt{de Jager} and \texttt{Yang} schemes is about 1 to 2\,$M_{\rm \odot}$ at the same log\,$L_{\rm prog}$. 

Despite the variation among different wind schemes, each of these schemes predicts a unique $M_{\rm Henv}$ given the same log\,$L_{\rm prog}$ without introducing additional scatter. This cannot explain the large scatter in the log\,$L_{\rm prog}$-$M_{\rm Henv}$ relation seen in SNe II. A possible explanation is that, even for the same physical parameters, such as $T_{\rm eff}$ and log\,$L_{\rm prog}$, the mass-loss rate $\dot{M}$ of RSG is not uniquely determined. Indeed, the wind prescriptions adopted in the work (and most stellar evolution codes) are empirically derived from the $average$ $\dot{M}$ of RSGs with the same physical parameters. For example, for \texttt{Yang} scheme, $\dot{M}_{\rm Y23}$ has an intrinsic scatter of 0.5 dex (equivalent to a factor of $\sim$3 in linear scale) given the same log\,$L_{\rm prog}$. \citet{yang23} suggest a similar scatter level should be applied to all wind prescriptions (see also \citealt{mauron11}).  

\begin{figure*}
\epsscale{1}
\plotone{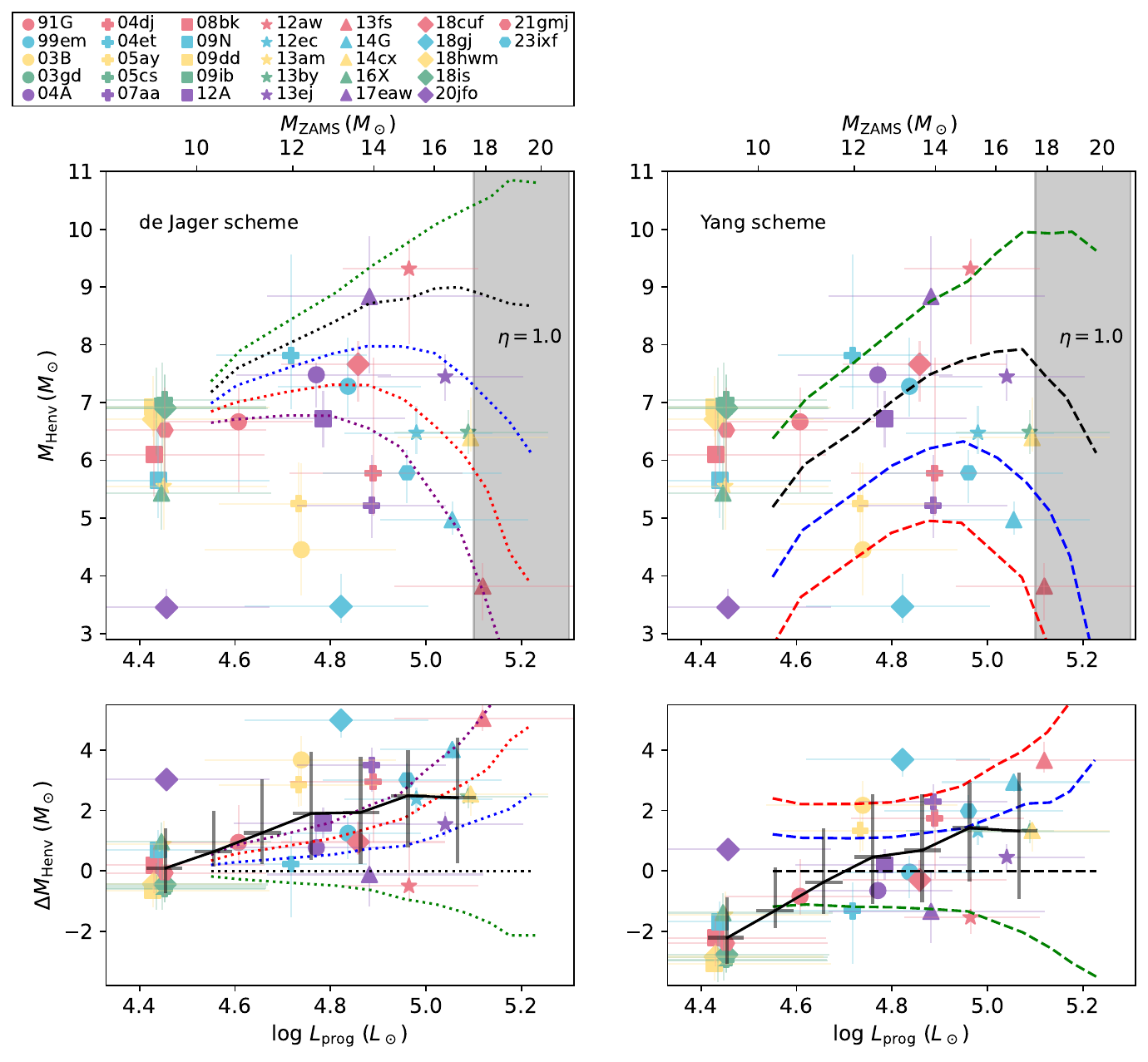}
\centering
\caption{Upper panels: same as Figure~\ref{fig:main}, with the log\,$L_{\rm prog}$-$M_{\rm Henv}$ of different wind prescriptions overlaid for comparison. The colored lines represent the predictions with wind efficiency $\eta$\,=\,0.5 (green), 1.0 (black), 1.5 (blue), 2.0 (red) and 2.5 (purple). The scatter points are the measurements for individual SNe in the sample, represented by the same color and marker as Figure~\ref{fig:main}. Lower panels: log\,$L_{\rm prog}$ versus the additional mass-loss $\Delta M_{\rm Henv}$. The black solid line is the average log\,$L_{\rm prog}$-$\Delta M_{\rm Henv}$ of the observed SNe, and the transparent error bars represent the 68\% CI. Left panels: \texttt{de Jager} scheme; Right panels: \texttt{Yang} scheme.}
\label{fig:effect_wind}
\end{figure*}

To investigate how the intrinsic scatter in mass-loss rates will affect the light curves of SNe II, we evolve additional model grids identical to those in \S2.2.1, except that the wind mass-loss rates are artificially scaled by a constant factor:
\[
\dot{M}\,=\,\eta\,\dot{M}_{\rm J88} ({\rm or}~\dot{M}_{\rm Y23}),
\]
where $\eta$\,=\,[0.5, 1.5, 2]. For \texttt{de Jager} scheme, we additional evolve a grid with $\eta$\,=\,2.5. The results are shown in Figure~\ref{fig:effect_wind}, where the models are compared with the log\,$L_{\rm prog}$-$M_{\rm Henv}$ of SNe II. In the lower panels, the additional mass-loss $\Delta M_{\rm Henv}$ are shown as function of log\,$L_{\rm prog}$. Here $\Delta M_{\rm Henv}$ is defined as the difference between $M_{\rm Henv}^{\rm S}$ and $M_{\rm Henv}$, where $M_{\rm Henv}^{\rm S}$ represents the envelope mass of the models with $\eta$\,=\,1.0 at the same log\,$L_{\rm prog}$. The average of the log\,$L_{\rm prog}$-$\Delta M_{\rm Henv}$ of SNe II are shown as the black solid line, and the 68\% CI is represented by the error bars.

For both schemes, $M_{\rm Henv}$ is strongly affected by $\eta$ at the bright end. For the \texttt{de Jager} scheme, varying $\eta$ by a factor of 2 will only lead to a variation of \(<\)\,1.0\,$M_{\rm \odot}$ in $M_{\rm Henv}$ at log\,$L_{\rm prog}$\,\(<\)\,4.8, and cannot account for the observed diversity of $M_{\rm Henv}$ at this range. However, \texttt{Yang} scheme can create RSG progenitors with low $M_{\rm Henv}$ at the faint end with $\eta$\,=\,1.5 (0.18 dex), within the scatter level proposed by \citet{yang23}. With $\eta$\,$\in$\,[0.5, 2.0], \texttt{Yang} scheme can well cover the log\,$L_{\rm prog}$-$M_{\rm Henv}$ scatter of SNe II.

However, in the comparison of log\,$L_{\rm prog}$-$\Delta M_{\rm Henv}$, the results of \texttt{Yang} scheme shows a less satisfactory agreement: observations suggest an increasing trend in the log\,$L_{\rm prog}$-$\Delta M_{\rm Henv}$ relation for SNe II, whereas the \texttt{Yang} scheme predicts a much flatter relation. To match the observed trend, the \texttt{Yang} scheme would require damped mass loss ($\eta$\,\(<\)\,1.0) for faint RSGs and enhanced mass loss (by a factor of 1.5) for bright RSGs. Such a luminosity-dependent $\eta$ is unexpected if the observed scatter in log\,$L_{\rm prog}$-$M_{\rm Henv}$ arises $solely$ from the intrinsic scatter in $\dot{M}_{\rm Y23}$; otherwise, the observed data points would be expected to scatter around the zero line without clear dependence of  $\eta$ on log\,$L_{\rm prog}$. 

In conclusion, the \texttt{Yang} scheme, when incorporating the intrinsic scatter in $\dot{M}_{\rm Y23}$, provides a better explanation for the observed scatter in the log\,$L_{\rm prog}$-$M_{\rm Henv}$ relation of SNe II compared to the commonly adopted \texttt{de Jager} scheme. However, it fails to reproduce the observed log\,$L_{\rm prog}$-$\Delta M_{\rm Henv}$ relation. This discrepancy may stem from the empirical nature of $\dot{M}_{\rm Y23}$, which is derived from a large sample of M-type RSGs, most of which are in long-period states with relatively weak winds. If mass loss occurs in an eruptive manner, where a significant fraction of the envelope is ejected over short timescales, such events could be underrepresented by the current observations, making direct modeling of these processes challenging (\citealt{meynet15}). If these process are log\,$L_{\rm prog}$ dependent, such that more luminous RSGs are more likely to expel their hydrogen-rich envelope through these eruptive processes which are currently unaccounted in the \texttt{Yang} scheme (see for example, \citealt{yoon_pulse}), it may potentially explain the dependence of $\eta$ on log\,$L_{\rm prog}$.

Given the uncertainty in the cool wind schemes, and the relatively large intrinsic scatter in $\dot{M}$ within a fixed scheme, we propose that, even if only single stars are considered as the progenitor of SNe II, as assumed for most studies, $M_{\rm Henv}$ is better treated as a free parameter (this work; see also \citealt{hiramatsu21,ravi24,fang25a,fang25b}) rather than as a quantity strictly determined by log\,$L_{\rm prog}$ (or $M_{\rm ZAMS}$). In \S5.1, we will demonstrate how imposing a fixed relation between $M_{\rm Henv}$ and $M_{\rm ZAMS}$ can introduce biases in parameter inference from light curve modeling.

\subsection{Binary interaction}
In the previous section, we have discussed the effect of different wind mass schemes on the log\,$L_{\rm prog}$-$M_{\rm Henv}$ relation, assuming SNe II originating from RSGs evolved from single stars. However, observations indicate that most massive stars are born in binary systems (see, e.g., \citealt{sana12}), where binary interactions can strip the hydrogen-rich envelope to varying degrees. In this section, we explore whether, given the distribution of initial orbital parameters, binary evolution can account for the observed scatter in the log\,$L_{\rm prog}$-$M_{\rm Henv}$ relation from a population perspective, using the binary model grid evolved in \S2.2.2.

\subsubsection{Properties of the binary grid}
\begin{figure*}
\epsscale{1.1}
\plotone{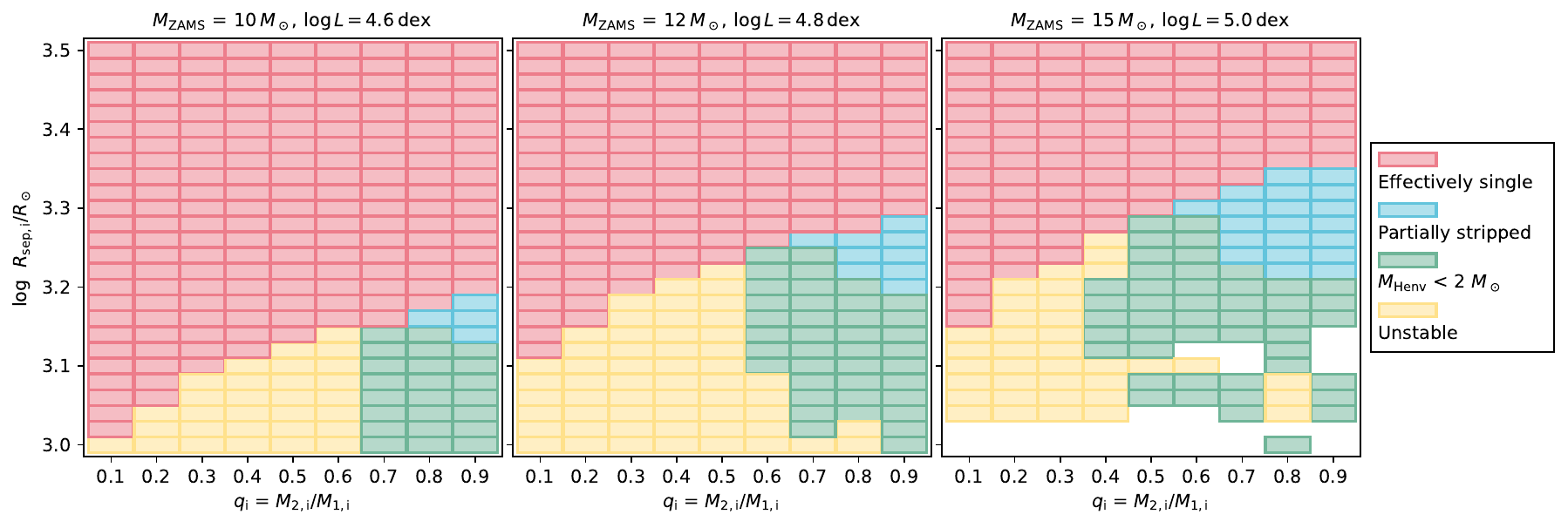}
\centering
\caption{Initial mass ratio ($q_{\rm i}$) versus initial separation (log\,$R_{\rm sep,i}$) for binary evolution models with a fixed initial primary mass (represented by its $M_{\rm ZAMS}$). Different colors indicate distinct evolutionary outcomes, while unfilled regions represent non-convergent cases. Each panel corresponds to a different primary $M_{\rm ZAMS}$. Left: $M_{\rm ZAMS}$\,=\,10\,$M_{\rm \odot}$; Middle: $M_{\rm ZAMS}$\,=\,12\,$M_{\rm \odot}$; Right: $M_{\rm ZAMS}$\,=\,15\,$M_{\rm \odot}$.}
\label{fig:binary_grid}
\end{figure*}

The outcomes of the binary models are classified into 4 categories:
\begin{itemize}
    \item {\bf Effective single star}. If the initial separation is large enough, the primary and the companion stars will not have any mass transfer throughout the evolution and evolve independently as if they were single stars.
    \item {\bf Partially stripped}. For these systems, the initial separation is close enough to allow for binary mass transfer, and the residual $M_{\rm Henv}$ of the primary star falls within the range
    \[
        2\,M_{\rm \odot}\,<\,M_{\rm Henv}\,<\,M_{\rm Henv}^{\rm S} - 1\,M_{\rm \odot},
    \]
    where $M_{\rm Henv}^{\rm S}$ is the hydrogen-rich envelope mass if it was evolved as a single star.
    \item {\bf Hydrogen deficient}. Similar to the above case, but with 
    \[
        M_{\rm Henv}\,<\,2\,M_{\rm \odot}.
    \]
    These stars would not explode as SNe II in our sample because the $M_{\rm Henv}$ is too small.
    \item {\bf Unstable mass transfer}.
\end{itemize}
The ranges of $q_{\rm i}$ and $R_{\rm sep,i}$ corresponding to these categories are indicated by the differently colored regions in Figure~\ref{fig:binary_grid}. There are also cases where the companion star expands during mass accretion, leading to a convergent problem (represented by the unfilled regions in Figure~\ref{fig:binary_grid}; see \citealt{lau24}). However, these situations only occur when the initial separation of the two stars is so small that they fall into the hydrogen-deficient or unstable mass-transfer categories, which are irrelevant with this work and thus neglected.

The results are generally consistent with \citet{ercolino23} for fixed primary $M_{\rm ZAMS}$ and can be summarized as follows: (1) the boundary of log\,$R_{\rm sep,i}$ that separates systems with and without binary mass transfer increases with $q_{\rm i}$; (2) for systems with low $q_{\rm i}$, binary mass transfer tends to be unstable once it occurs (columns with only pink and yellow regions in Figure~\ref{fig:binary_grid}); (3) mass loss through binary interaction behaves as an 'all-or-nothing' process: once mass transfer occurs, it typically removes nearly the entire hydrogen-rich envelope. Consequently, it is difficult to produce partially stripped primaries, as reflected in the limited extent of the light blue regions in Figure~\ref{fig:binary_grid}. Partial stripped primary is only feasible for systems with relatively large $q_{\rm i}$ and a narrow range of $R_{\rm sep,i}$. Since the main focus of this work is on the formation channel of partially stripped RSGs, the following discussion mainly addresses the implications of feature (3).

The 'all-or-nothing' nature of hydrogen-rich envelope stripping through binary interaction stems from the convective structure of the envelope, which has a flat or negative entropy gradient (i.e., lower entropy at the surface). When the envelope expands to fill the Roche lobe, mass transfer occurs, reducing the envelope mass and radius. As the mass and radius decreases, higher-entropy regions are exposed, causing the envelope to re-expand to its original size on a dynamical timescale, filling the Roche lobe and triggering binary interaction again. This process continues until the radiative layer at the bottom $\sim$1\,$M_{\rm \odot}$ of the envelope, which is characterized by a positive entropy gradient, is reached. At this point, the envelope cannot re-expand to fill the Roche lobe because the surface now has lower entropy than its previous state, halting the mass transfer process (see also discussion in \citealt{soberman97,ivanova11a,ivanova11b,hirai22}).

One possible channel for forming partially stripped stars is if the primary star explodes midway through mass loss, i.e., while binary mass transfer is still ongoing at the time of explosion. Indeed, all partially stripped models in our grid (light blue regions in Figure~\ref{fig:binary_grid}) show high mass-loss rates (10$^{-4}$ to 10$^{-3}$\,$M_{\rm \odot}\,{\rm yr}^{-1}$) due to binary interaction at the endpoint of the calculation. This also explains why partially stripped models are more likely for systems with higher primary $M_{\rm ZAMS}$: stars with larger $M_{\rm ZAMS}$ evolve more quickly, being more likely to explode before the hydrogen-rich envelope is completely stripped.

\subsubsection{Population properties}
\begin{figure}
\epsscale{1.}
\plotone{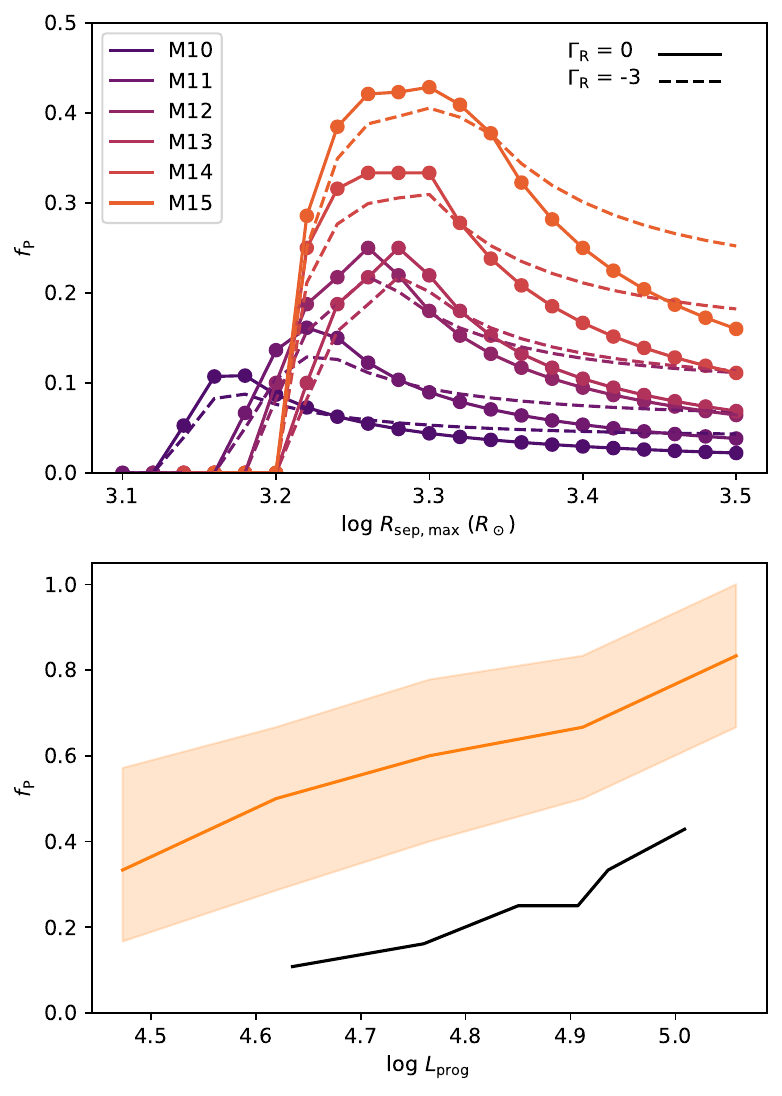}
\centering
\caption{Upper panel: Fraction of binary systems in which the primary star evolves into a partially stripped RSG by the end of the calculation, shown as function of log\,$R_{\rm sep,max}$. Different colors correspond to different initial primary masses. Solid lines represent cases where log\,$R_{\rm sep,i}$ and $q_{\rm i}$ follow flat distributions, while dashed lines correspond to steeper distributions with $\Gamma_{R}\,=\,-3$ for log\,$R_{\rm sep,i}$ (Equation~\ref{eq:r_distribution}). Lower panel: Fraction of observed SNe II originating from partially stripped RSGs as a function of log\,$L_{\rm prog}$. The shaded region represents the 68\% CI. The black solid line indicates the maximum possible fraction of partially stripped RSGs produced through binary interactions (corresponding to the peaks of the solid lines in the upper panel) as a function of log\,$L_{\rm prog}$.}
\label{fig:binary_fp}
\end{figure}

Our next step is to investigate whether binary interaction can explain the population of partially stripped RSGs expected for SNe II progenitors. We start from a simple assumption, i.e., flat distributions for $q_{\rm i}$ and log\,$R_{\rm sep,i}$, which is broadly consistent with observation (see, e.g., \citealt{sana12,villasenor25}). We also assume upper cutoff for log\,$R_{\rm sep,max}$, i.e., binaries can only be born with initial separation within this range. The problem then reduces to counting the light blue and pink areas (denoted as $N_{\rm P}$ and $N_{\rm S}$ respectively; 'P' refers to partially stripped and 'S' refers to effectively single) with log\,$R_{\rm sep,i}$\,\(<\)\,log\,$R_{\rm sep,max}$. The fraction of partially stripped RSGs, defined as 
\begin{equation}
    f_{\rm P}\,=\,\frac{N_{\rm P}}{N_{\rm P}\,+\,N_{\rm S}},
\end{equation}
is shown in Figure~\ref{fig:binary_fp} as function of log\,$R_{\rm sep,max}$.

For a fixed $M_{\rm ZAMS}$, the fraction of partially stripped RSGs produced through binary interaction strongly depends on the upper cutoff of the initial separation. Taking the M12 model as an example: initially, when log\,$R_{\rm sep,max}$\,\(<\)\,3.16, no partially stripped primaries are formed and $f_{\rm P}$\,=\,0. As log\,$R_{\rm sep,max}$ increases, partially stripped primaries begin to appear for $q_{\rm i}$\,\(>\)\,0.7, and $f_{\rm p}$ rises accordingly, reaching a maximum $f_{\rm P}$\,=\,0.25 at log\,$R_{\rm sep,max}$\,=\,3.28. Beyond this value, binary interaction can no longer create partially stripped RSGs, and further increases in log\,$R_{\rm sep,max}$ only add to the effective single category ($N_{\rm S}$), reducing $f_{\rm p}$. If all RSGs with $M_{\rm ZAMS}$\,=\,12\,$M_{\rm \odot}$ were born in binary systems as primary stars, at most 25\% of them could be partially stripped.

The above analysis is based on the assumption that both $q_{\rm i}$ and log\,$R_{\rm sep,i}$ follow flat distributions. However, in reality, the distribution of log\,$R_{\rm sep,i}$ is more complicated (\citealt{sana12,moe17,moe19}). To investigate how this affects the result, we model the distribution of log\,$R_{\rm sep,i}$ using a power-law function:
\begin{equation}
    \frac{dN}{d\,{\rm log}\,R_{\rm sep,i}}\,\propto\,R_{\rm sep,i}^{\Gamma_{\rm R}},
\label{eq:r_distribution}
\end{equation}
where a flat distribution in log\,$R_{\rm sep,i}$ corresponds to $\Gamma_{\rm R}$\,=\,0.

Assuming that $q_{\rm i}$ still follows a flat distribution, we repeat the analysis with $\Gamma_{\rm R}$\,=\,-3. The results, shown as dashed lines in the upper panel of Figure~\ref{fig:binary_fp}, indicate that the maximum $f_{\rm p}$ is consistently lower compared to the flat distribution cases ($\Gamma_{\rm R}$\,=\,0). Our experiments with different $\Gamma_{R}$ values show that, for fixed $M_{\rm ZAMS}$, the maximum $f_{\rm p}$ is a decreasing function of $\Gamma_{\rm R}$. The peak of the solid line in the upper panel of Figure~\ref{fig:binary_fp} therefore represents the maximum possible fraction of partially stripped primaries produced by binary interaction for a fixed $M_{\rm ZAMS}$, as long as close pairs are strongly favored over a flat distribution in log\,$R_{\rm sep,i}$ ($\Gamma_{\rm R}$\,\(<\)\,0; \citealt{sana12,moe17,moe19}). For M12 models, this maximum value is 0.25, while for M15 models, it increases to 0.42.

The maximum $f_{\rm p}$, as function of log\,$L_{\rm prog}$, is shown as the black solid line in the lower panel of Figure~\ref{fig:binary_fp}. The observed log\,$L_{\rm prog}$-$f_{\rm P}$ relation is constructed as follow:
\begin{itemize}
    \item For each SNe in the sample, we randomly pick $l$ from its posterior log\,$L_{\rm prog}$ distribution, and the corresponding envelope mass $m$ is estimated following the procedure outlined in \S3;
    \item The expected $M_{\rm Henv}^{\rm S}$ at $l$ for single star model is estimated from the log\,$L_{\rm prog}$-$M_{\rm Henv}$ relation of the \texttt{KEPLER} models. If $m$\,\(<\)\,$M_{\rm Henv}^{\rm S}$\,-\,1\,$M_{\rm \odot}$, it is labeled as 'partially-stripped';
    \item The random sample is binned into 6 groups according to the order of $l$ (averagely 5 objects in each), and we compute the fraction of partially-stripped objects in each bin;
    \item The processes are repeated for 10,000 times. The median values of $f_{\rm P}$ in each log\,$L_{\rm prog}$ bin, along with the 68\% CI, are represented by the orange solid line and the shaded region in the lower panel of Figure~\ref{fig:binary_fp}.
\end{itemize}

Although the observed log\,$L_{\rm prog}$-$f_{\rm P}$ relation shows an increasing trend, the SNe II sample is biased toward well-observed objects, and the CI range is too large to support a statistically meaningful conclusion. But still, the data clearly indicate that partially-stripped RSG, as progenitors of SNe II, are not rare across the log\,$L_{\rm prog}$ range. For log\,$L_{\rm prog}$\,=\,4.5, $f_{\rm P}$\,=\,0.42$^{+0.25}_{-0.19}$, which further increases to 0.75$^{+0.18}_{-0.17}$ at log\,$L_{\rm prog}$\,=\,5.0. Given the same log\,$L_{\rm prog}$, the maximum $f_{\rm P}$ predicted by binary models falls out of the 2$\sigma$ range of the observed one. 

In conclusion, while binary interaction is an efficient mechanism for removing the hydrogen-rich envelope of the primary star and may explain the light-curve properties of several individual SNe II (see, e.g., \citealt{dessart24,michel25}), it is too efficient to account for the observed fraction of partially stripped RSGs in the overall population. However, the $f_{\rm p}$ predicted by our binary models still falls short in capturing the full complexity of the binary population:
\begin{enumerate}[label=(\arabic*)]
    \item The maximum $f_{\rm p}$ for fixed log\,$L_{\rm prog}$ ($M_{\rm ZAMS}$) is adopted to compare with observation. However, taking the M12 models as example, this corresponds to a log\,$R_{\rm sep,i}$ cutoff at log\,$R_{\rm sep,max}$\,=\,3.28, while observation indicates that log\,$R_{\rm sep,i}$ can extend to larger values (\citealt{sana12, moe17, moe19});
    \item We do not account for the evolution of the companion star, which may also undergo core-collapse if its mass exceeds $\sim$\,8\,$M_{\rm \odot}$. The companion star will evolve as a single star or in a binary system with the compact remnant left by the explosion of the original primary star with $M_{\rm compact}$\,$\sim$\,1.4\,$M_{\rm \odot}$, depending on whether the system remains bounded after the explosion. In the latter case, the system has $q_{\rm i}$\,\(<\)\,1.4/8.0\,$\sim$\,0.18, and the mass transfer is very unstable (Figure~\ref{fig:binary_grid}). The companion will eventually fall into the unstable or effective single category. In the latter case, the value of $f_{\rm P}$ decreases further;
    \item In the models of \citet{hirai20}, if the final separation of the binary system is close enough, the explosion of the primary star can help to remove the hydrogen-rich envelope of the companion star. Their two-dimensional simulations suggest that approximately 50\% to 90\% of the hydrogen-rich envelope of the companion star may become unbound through this process, potentially leading to the formation of partially stripped RSGs.
\end{enumerate}
With these factors taken into account, the tension between the observed $f_{\rm P}$ and the prediction from binary models is either relaxed or worsened, depending on the relative occurrence rates of these channels.

The conclusion above is based on binary models that undergo stable mass transfer. However, systems experiencing unstable mass transfer, which constitute the majority of binary populations (\citealt{sana12}), should not be neglected. For instance, in the M10 binary grid, assuming log\,$R_{\rm sep,i}$ is uniformly distributed within [0, 3.2] (since systems with log\,$R_{\rm sep,i}$\,\(>\)\,3.2 effectively behave as single stars), at least 75\% of the binary systems will end up as unstable mass transfer. These systems are likely to experience common envelope evolution, during which the companion spirals into the hydrogen-rich envelope of the RSG primary, potentially leading to efficient mass ejection (see, e.g., \citealt{klencki21}). However, this process is highly non-linear, and only a limited number of multi-dimensional simulations are currently available (see, e.g., \citealt{lawsmith20,lau22,moreno22,CE_review,shiber24,vetter24,lau25}). The outcome of the merger and its fate are also uncertain (see e.g. \citealt{ivanova13} for a review; \citealt{schneider21,renzo23,schneider24}): the spiral in of the companion can mix the entire envelope (see, e.g., \citealt{ivanova02}), and the post-merger envelope becomes completely homogeneous, leading to the formation of compact blue supergiant (BSG) which explodes as SN 1987-like events \citep{menon17,urushibata18,menon19,utrobin21,menon24,moriya24}. Depending on the degrees of mixing, the star may also re-expand as RSG. However, the structures of these post-merger RSGs, characterized by massive, highly mixed hydrogen-rich envelope supported by a relatively low mass He core, may be very different from their counterparts evolved as non-rotating single stars as assumed in this work (see e.g. \citealt{farrell20b}). The merger products should be further investigated in the future to assess whether they can account for the observed fraction of partially-stripped RSGs inferred from plateau phase light curves of SNe II. 

\subsection{More compact RSG progenitor?}
\begin{figure*}
\epsscale{1.1}
\plotone{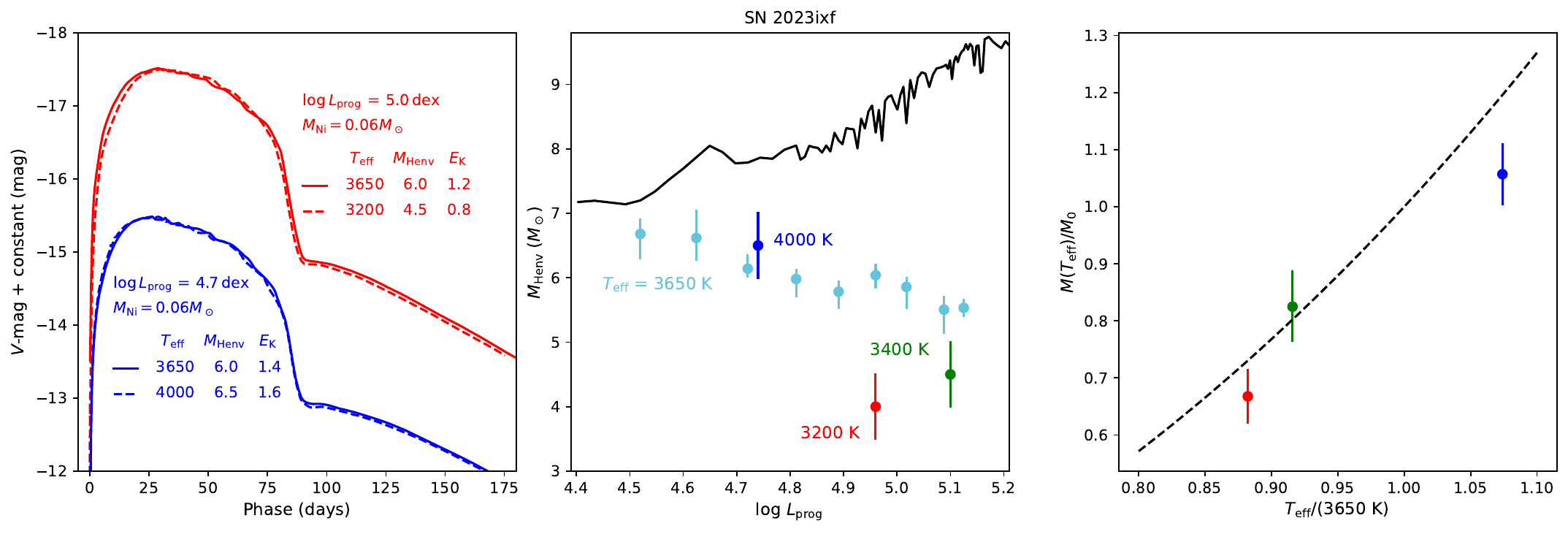}
\centering
\caption{The effect of RSG progenitors $T_{\rm eff}$ on light curve modeling of SN 2023ixf. Left panel: Optimized $V$-band light curves for SN 2023ixf with different RSG progenitor properties. Red solid line: log\,$L_{\rm prog}$\,=\,5.0 and $T_{\rm eff}$\,=\,3650\,K (this work); Blue solid line: log\,$L_{\rm prog}$\,=\,4.7 and $T_{\rm eff}$\,=\,3650\,K (this work); Red dashed line: log\,$L_{\rm prog}$\,=\,5.0 and $T_{\rm eff}$\,=\,3200\,K (\citealt{vandyk24,fang25b}); Blue dashed line: log\,$L_{\rm prog}$\,=\,4.7 and $T_{\rm eff}$\,=\,4000\,K (\citealt{kilpatrick23,fang25b}). Middle panel: Inferred $M_{\rm Henv}$ from light curve modeling of SN 2023ixf as a function of log\,$L_{\rm prog}$. Light blue scatter points denote measurements from this work (progenitor RSGs with $T_{\rm eff}\,\sim\,3650\,$K). Red, green and blue points are the measurements in \citet{fang25b} for progenitor RSGs with different $T_{\rm eff}$. The black solid line is the log\,$L_{\rm prog}$-$M_{\rm Henv}$ relation of \texttt{KEPLER} models. Right panel: The ratio of the inferred $M_{\rm Henv}$ between RSG models with effective temperature $T_{\rm eff}$ and the measurement of this work, given the same log\,$L_{\rm prog}$, plotted as a function of $T_{\rm eff}$/3650\,K. The dashed line represents Equation~\ref{eq:teff_relation}.}
\label{fig:compactness}
\end{figure*}

In this work, we assume the progenitors of SNe II are RSGs with $T_{\rm eff}$ ranging from 3500 to 3700\,K, similar to the values of field RSGs. However, this assumption does not necessarily hold: massive stars form at sites that are expected to correlate with H-II regions, spanning a wide range of metallicity (\citealt{dessart14,taddia16,anderson16,2015bs,gutierrez18,scott19,tucker24}). The variations in the physical conditions of the progenitor's birth place will affect the properties of the RSG. For example, our own experiment with \texttt{MESA} shows that, if all the other parameters are the same, RSGs with low metallicity will tend to be more compact with higher $T_{\rm eff}$ (see also \citealt{dessart13,dessart14}). Hereafter, we use $T_{\rm eff}$ rather than the radius $R_{\rm prog}$ as the indicator of the envelope compactness of the RSG progenitors. For clarity, 'compactness' here always refers to the radius of the entire star, not to the inner-core compactness parameter often invoked in discussions of core-collapse mechanism. While $R_{\rm prog}$ characterizes the size of an individual RSG, $T_{\rm eff}$ serves as a more representative measure of the envelope compactness of an RSG population.

The envelope compactness of the RSG can affect the inferred $M_{\rm Henv}$ from plateau-phase light curve modeling. For a fixed log\,$L_{\rm prog}$ (measured from nebular-phase spectroscopy), more compact RSG progenitors have smaller radii. As a result, their explosions require higher $E_{\rm K}$ to reach the same plateau-phase magnitude, which shortens the plateau duration. To compensate for this reduction, a larger $M_{\rm Henv}$ is needed. 

To illustrate these effects, we compare the light curve models from RSG progenitors with solar metallicity and $T_{\rm eff}$ = 3200 K ($\alpha_{\rm MLT}=1.8$) and 3900 K ($\alpha_{\rm MLT}=2.8$), using \{$M_{\rm Henv}$, $E_{\rm K}$, $M_{\rm Ni}$\} values that best fit the light curve of SN 2023ixf (\citealt{fang25b}), with the optimized light curve for the same object from this work. Taking the M12 models (log\,$L_{\rm prog}$ = 4.7) as an example: in \citealt{fang25b}, using RSG models with $T_{\rm eff}$ =\,3900\,K, $M_{\rm Henv}$ of SN 2023ixf is estimated to be 6.5\,$M_{\rm \odot}$, slightly larger than the 6.0\,$M_{\rm \odot}$ inferred in this work ($T_{\rm eff}$\,=\,3650\,K). However, these two light curves show no visible difference (left panel of Figure~\ref{fig:compactness}). This effect becomes more pronounced for the M15 models, where the difference in $M_{\rm Henv}$ can be as large as 1.5\,$M_{\rm \odot}$. In the middle panel of Figure~\ref{fig:compactness}, we compare the inferred $M_{\rm Henv}$ as a function of log\,$L_{\rm prog}$ for models with different $T_{\rm eff}$, revealing a clear trend: given the same log\,$L_{\rm prog}$, more compact RSGs (with higher $T_{\rm eff}$) yield larger $M_{\rm Henv}$ estimates, and this dependency can be described in power-law form:
\begin{equation}
    \frac{M{\rm (}T_{\rm eff}{\rm )}}{M_{\rm 0}}\,=\,(\frac{T_{\rm eff}}{\rm 3650\,K})^{2.52},
\label{eq:teff_relation}
\end{equation}
as illustrated in the right panel of Figure~\ref{fig:compactness}. Here $M{\rm (}T_{\rm eff}{\rm )}$ is $M_{\rm Henv}$ estimated using the light curves from RSG progenitor models with effective temperature $T_{\rm eff}$, and $M_{\rm 0}$ is the estimation in this work ($T_{\rm eff}$\,=\,3650\,K), with the same log\,$L_{\rm prog}$ ($M_{\rm ZAMS}$). This relation is similar to the degeneracy curve with respect to $R_{\rm prog}$ introduced in \citet{goldberg20}.

The high fraction of partially stripped RSGs proposed as progenitors of SNe II is based on light curve modeling with progenitor RSGs having $T_{\rm eff}$\,$\sim$\,3650\,K. From Equation~\ref{eq:teff_relation}, if the actual progenitor RSGs are more compact, the inferred $M_{\rm Henv}$ would increase accordingly. In that case, there would be no need to invoke partially stripped RSGs as the progenitors of SNe II. The expected $T_{\rm eff}$ is estimated as
    \[
        T_{\rm eff}\,=\,{\rm 3650}\,(\frac{M_{\rm Henv}^{\rm S}}{M_{\rm Henv}})^{\frac{1}{2.52}}\,{\rm K},
    \]
making use of Equation~\ref{eq:teff_relation}. Here $M_{\rm Henv}^{\rm S}$ is the prediction of the \texttt{KEPLER} models at the same log\,$L_{\rm prog}$, and $M_{\rm Henv}$ is the log\,$L_{\rm prog}$-weighted value.

\begin{figure}
\epsscale{1.}
\plotone{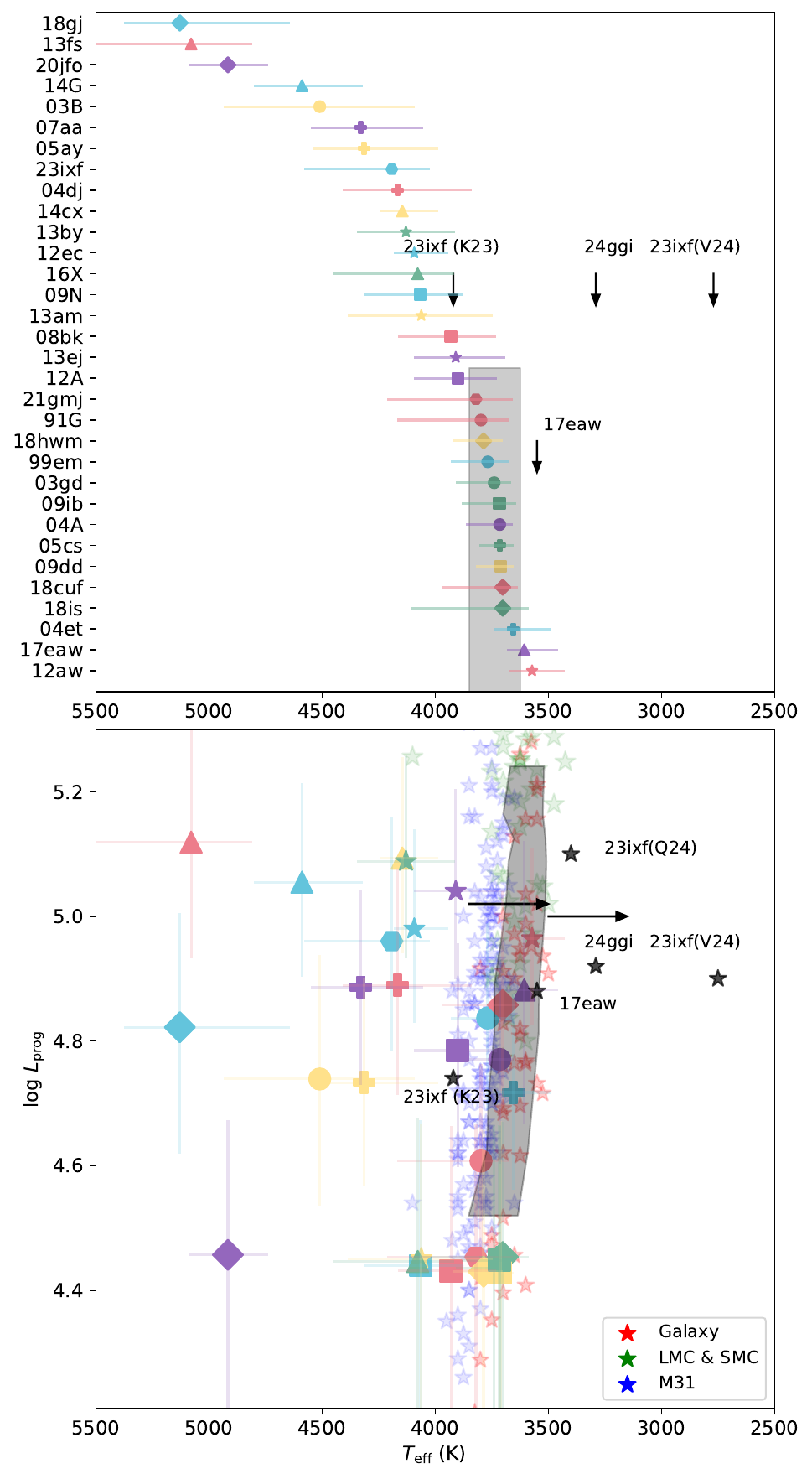}
\centering
\caption{Upper panel: Cumulative distribution of the progenitor $T_{\rm eff}$ for SNe II in the sample, assuming their $M_{\rm Henv}$ follows predictions from \texttt{KEPLER} models. The shaded region represents the 68\% CI for the $T_{\rm eff}$ of M-type RSGs observed in the field. Vertical arrows indicate $T_{\rm eff}$ values inferred from pre-SN images for several well-observed SNe II. Lower panel: SNe II progenitors plotted on the HRD, where log\,$L_{\rm prog}$ is inferred from nebular spectroscopy, and $T_{\rm eff}$ corresponds to the values shown in the upper panel. Transparent colored stars represent M-type RSGs in the field, while black stars denote RSG progenitors of well-observed SNe II identified from pre-SN images. The black strip marks the progenitor RSG models used in this work (Table~\ref{tab:model_grid}). Horizontal arrows indicate a scale of 200\,K.}
\label{fig:teff_distribution}
\end{figure}

The distribution of $T_{\rm eff}$ estimated in this way is shown in the upper panel Figure~\ref{fig:teff_distribution}. The shaded region represents the 68\% CI for $T_{\rm eff}$ values of M-type RSGs in the field, including those observed in the Galaxy (\citealt{levesque05}), M31 (\citealt{massey16}), Large Magellanic Cloud (LMC; \citealt{levesque06}), and Small Magellanic Cloud (SMC; \citealt{levesque06}). Approximately 15 SNe have expected $T_{\rm eff}$ values that fall in this range within 1$\sigma$. For these objects, there is no need to invoke partially stripped RSGs as their progenitors.

However, for the remaining SNe, the required $T_{\rm eff}$ values are significantly higher than those of field RSGs. For example, if SN 2018gj originated from the explosion of an RSG with a hydrogen-rich envelope predicted by \texttt{KEPLER} models at the same log\,$L_{\rm prog}$, its expected $T_{\rm eff}$ would be 5128$^{+242}_{-480}$\,K to explain the relative short plateau light curve. In the lower panel of  Figure~\ref{fig:teff_distribution}, the positions of the RSG progenitors of SNe II in the sample are shown in the Hertzsprung-Russell Diagram (HRD). The log\,$L_{\rm prog}$ is inferred from the nebular phase spectroscopy using the observation-calibrated MLR (\citealt{fang25c}; see also Table~\ref{tab:sample_appendix}), and $T_{\rm eff}$ is constrained by the plateau phase light curve modeling, if the progenitors follow the same log\,$L_{\rm prog}$-$M_{\rm Henv}$ relation of the \texttt{KEPLER} models. 

This comparison suggests that, if we abandon the assumption of partial stripping, SNe II would arise from a population of RSGs that is different from those observed in the field. In fact, the intrinsic tension might be more severe: (1)\,Compact RSGs are usually born in low metallicity environment (see, for example, \citealt{dessart13}), and have smaller radii. As a result, the stellar wind is weaker than the ones employed in \texttt{KEPLER}, and $M_{\rm Henv}^{\rm S}$ from these models should be regarded as lower estimation. The estimated $T_{\rm eff}$ is expected to be even higher than the current values; (2)\,Most field RSGs are in the helium-burning phase, as the timescale for helium burning is significantly longer than those of later evolutionary stages. Our own experiment with \texttt{MESA} shows that, at carbon burning phase, the hydrogen-envelope will expand, reducing $T_{\rm eff}$ at core carbon depletion by 100 to 300\,K compared to its value at core helium depletion. The field RSGs in the lower panels of Figure~\ref{fig:teff_distribution} should shift to cooler side (right) by this amount at the onset of the explosion, further increasing the aforementioned discrepancy.

The assumption of more compact progenitor is also not supported by the $T_{\rm eff}$ values of SNe II with pre-SN images. 
In the lower panel of Figure~\ref{fig:teff_distribution}, the HRD positions of the RSG progenitors for 3 well-observed SNe II are plotted for comparison (see discussion in \citealt{beasor25} for the importance of multi-band photometry in pre-SN images): SN 2017eaw with $T_{\rm eff}$\,=\,3550\,K and log\,$L_{\rm prog}$\,=\,4.88 (\citealt{rui19_17eaw}), SN 2023ixf with $T_{\rm eff}$\,=\,2770\,K and log\,$L_{\rm prog}$\,=\,4.94 (\citealt{vandyk24}; V24), SN 2024ggi with $T_{\rm eff}$\,=\,3290\,K and log\,$L_{\rm prog}$\,=\,4.92 (\citealt{xiang24_ggi}). For SN 2023ixf, its progenitor's properties are quite uncertain (\citealt{pledger23,kilpatrick23, niu23, jencson23, neustadt24, xiang24, vandyk24, ransome24, qin23, soraisam23, liu23}; see also a summary in \citealt{fang25b}), and we plot two additional measurements for reference: $T_{\rm eff}$\,=\,3920\,K and log\,$L_{\rm prog}$\,=\,4.74 (\citealt{kilpatrick23}; K23); $T_{\rm eff}$\,=\,3400\,K and log\,$L_{\rm prog}$\,=\,5.10 (\citealt{qin23}; Q24)

Except for the measurement of K23, the progenitor RSGs of these SNe II appear cooler than field RSGs, consistent with our earlier discussion that $T_{\rm eff}$ decreases in the later stages of massive star evolution. In the lower panel of Figure~\ref{fig:teff_distribution}, the length of the black arrows represents a scale of 200\,K. If the $T_{\rm eff}$ of field RSGs were shifted downward by this amount, they could encompass the progenitor for SNe 2017eaw, 2023ixf (Q24), and 2024ggi, while the value from V24 remains too low to be explained by current models. Even for K23, the $T_{\rm eff}$ of 3920\,K is lower than the estimation for more than half of the SNe in our sample, if their progenitors follow the log\,$L_{\rm prog}$-$M_{\rm Henv}$ relation of the \texttt{KEPLER} models. 

In conclusion, although compact RSGs can explain light curve modeling results without invoking the assumption of partial stripping, the inferred $T_{\rm eff}$\,\(>\)\,4300\,K is not supported by any current observational evidence, including the $T_{\rm eff}$ range of field RSGs and the inferred $T_{\rm eff}$ values from pre-SN images for several well-observed SNe II.

\section{Discussions}
\subsection{Comparison with RSGs evolved with wind-driven mass-loss}
The core of the method in this work is to use nebular spectroscopy as additional constraint on the progenitor, which, however, is unavailable in many cases, limiting the applicability of this method to larger samples. Alternative strategies include introducing theoretical relations to reduce the dimensionality of the light curve modeling, such as the $M_{\rm ZAMS}$–$M_{\rm Henv}$–$R_{\rm prog}$ correlation derived from stellar evolution models (\citealt{morozova18, martinez22, moriya23, subrayan23}), or the $M_{\rm He\,core}$–$E_{\rm K}$ relation motivated by modern core-collapse theory \citep{curtis21,barker22,barker23}.

In this section, we focus on the former approach, which is commonly adopted in SNe II light curve modeling. Here, $M_{\rm ZAMS}$ is treated as the free parameter, while $R_{\rm prog}$ and $M_{\rm Henv}$ are uniquely determined by $M_{\rm ZAMS}$, reducing the system to three unknowns constrained by three observables. For convenience, we refer to these as 'wind models'. Although in \S4.1, we have shown that the selection of wind prescriptions, and the intrinsic scatter in the mass-loss rates will have an important effect on the final log\,$L_{\rm prog}$-$M_{\rm Henv}$ relation, to allow comparison with previous works, in this section, 'wind models' specifically refer to those that follow the $M_{\rm ZAMS}$-$M_{\rm Henv}$ relation of the KEPLER models, which are commonly used in SNe II light curve modeling. For the same reason, we also use $M_{\rm ZAMS}$ instead of log\,$L_{\rm prog}$ as the main parameter.

\begin{deluxetable*}{cccc|cccc}[t]
\centering
\label{tab:fit_zams}
\tablehead{
\colhead{SN}&\colhead{$M_{\rm ZAMS}$}&\colhead{$E_{\rm K}$}&\colhead{log\,$M_{\rm Ni}$}&\colhead{$M_{\rm ZAMS}$}&\colhead{$M_{\rm Henv}$}&\colhead{$E_{\rm K}$}&\colhead{log\,$M_{\rm Ni}$}
}
\startdata
2013by&10.86$_{-0.20}^{+0.14}$&1.463$_{-0.037}^{+0.024}$&-1.45$_{-0.03}^{+0.02}$&16.85$^{+3.15}_{-2.58}$&6.46$^{+0.39}_{-0.44}$&1.082$^{+0.135}_{-0.103}$&-1.36$^{+0.04}_{-0.07}$\\
2013fs&11.22$_{-0.22}^{+0.25}$&1.042$_{-0.038}^{+0.051}$&-1.31$_{-0.04}^{+0.04}$&17.72$^{+3.16}_{-2.83}$&3.82$^{+0.38}_{-0.58}$&0.527$^{+0.079}_{-0.085}$&-1.13$^{+0.03}_{-0.04}$\\
2014G&10.17$_{-0.13}^{+0.18}$&1.209$_{-0.049}^{+0.052}$&-1.37$_{-0.02}^{+0.01}$&16.28$^{+3.04}_{-2.17}$&4.97$^{+0.30}_{-0.25}$&0.656$^{+0.096}_{-0.067}$&-1.28$^{+0.02}_{-0.02}$
\enddata
\caption{The parameters \{$M_{\rm ZAMS}$, $E_{\rm K}$, log\,$M_{\rm Ni}$\} fitted with the wind models (left columns) compared with the $M_{\rm ZAMS}$ (converted from log\,$L_{\rm prog}$ derived from nebular spectroscopy using the MLR of the progenitor models in this work), and the log\,$L_{\rm prog}$-weighted \{$M_{\rm Henv}$, $E_{\rm K}$, log\,$M_{\rm Ni}$\} (right columns) for SNe 2013by, 2013fs and 2014G. $M_{\rm ZAMS}$, $M_{\rm Henv}$ and $M_{\rm Ni}$ are in the units of $M_{\rm \odot}$. $E_{\rm K}$ is in the unit of foe (10$^{51}$ erg).}
\end{deluxetable*}

\begin{figure*}
\epsscale{1.1}
\plotone{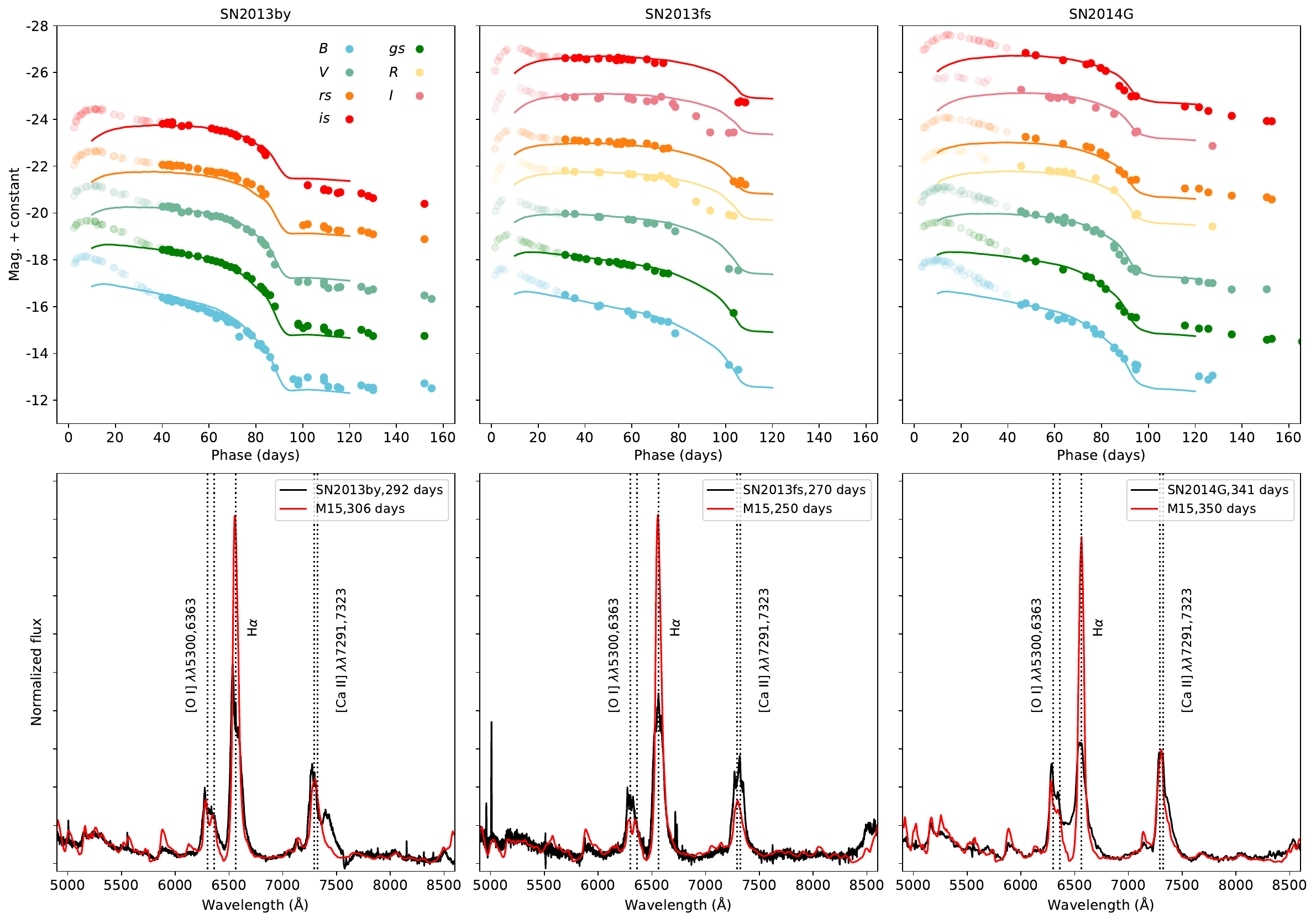}
\centering
\caption{Upper panels: The optimized multi-band light curves for SNe 2013by (left), 2013fs (middle) and 2014G (right), using the wind models. Lower panels: the nebular spectra of these 3 SNe (black), pre-processed and normalized following the procedure outlined in \citet{fang25c}, are compared with the nebular spectroscopy models from \citet{jerkstrand12} at similar phases (red). Compared to the M15 models, these objects show stronger [O I] lines but much weaker H$\alpha$. See also the similar comparison for SN 2023ixf in \citet{fang25b}.}
\label{fig:fit_wind}
\end{figure*}

We take 3 SNe, which have relative massive progenitor, inferred from the relatively strong [O I] seen in nebular spectroscopy, but low $M_{\rm Henv}$, estimated from light curve modeling, as examples: SNe 2013by, 2013fs and 2014G. These events serve as test cases to assess the validity of the $M_{\rm ZAMS}$ inferred from wind models. Similar to \S3, we use the \texttt{emcee} routine to infer \{$M_{\rm ZAMS}$, $E_{\rm K}$, log\,$M_{\rm Ni}$\} by fitting their plateau-phase light curves, assuming the same $M_{\rm ZAMS}$-$M_{\rm Henv}$ relation of the \texttt{KEPLER} models. For $M_{\rm ZAMS}$, we use a flat prior within the range of 10 to 18\,$M_{\rm \odot}$. The optimized parameters are shown in Table~\ref{tab:fit_zams}, compared with $M_{\rm ZAMS}$ transferred from log\,$L_{\rm prog}$ using the MLR of the models in this work, and the log\,$L_{\rm prog}$-weighted \{$M_{\rm Henv}$, $E_{\rm K}$, log\,$M_{\rm Ni}$\}. The resultant light curves are shown in the upper panels of Figure~\ref{fig:fit_wind}.

Even without prior knowledge on log\,$L_{\rm prog}$ ($M_{\rm ZAMS}$), we can still find reasonable fits to the observed light curves of SNe 2013by, 2013fs and 2014G, based on the wind models. However, the inferred $M_{\rm ZAMS}$ is significantly lower than the values independently constrained by nebular spectroscopy. For instance, log\,$L_{\rm prog}$ of SN 2013by is estimated to be 5.08$^{+0.16}_{-0.17}$ in \citet{fang25c}, which is converted to $M_{\rm ZAMS}$\,=\,16.85$^{+3.15}_{-2.58}$\,$M_{\rm \odot}$ using the MLR of the progenitor models in this work. In contrast, the wind model yields $M_{\rm ZAMS}$\,=\,10.86$_{-0.20}^{+0.14}$\,$M_{\rm \odot}$. The lower panels of Figure~\ref{fig:fit_wind} compare the nebular spectra of these three SNe with the spectral models from \citet{jerkstrand12}. All three SNe exhibit stronger [O I] and weaker H$\alpha$ than the M15 model at similar phases, suggesting they originate from massive RSGs with relatively low hydrogen-rich envelope masses, contrary to the lower progenitor masses inferred from the wind models.

This discrepancy arises from the dependence of the optimized $M_{\rm Henv}$ on $M_{\rm ZAMS}$. As shown in the upper left panel of Figure~\ref{fig:2014G_example}, for each fixed $M_{\rm ZAMS}$ (log\,$L_{\rm prog}$), we infer \{$M_{\rm Henv}$, $E_{\rm K}$, log\,$M_{\rm Ni}$\} using the \texttt{emcee} routine introduced in \S3. When $M_{\rm Henv}$ is a free parameter, its optimized value decreases with increasing $M_{\rm ZAMS}$. However, within our $M_{\rm ZAMS}$ range, the wind models predict an opposite trend (the black solid line in Figure~\ref{fig:main}). As a result, for the \texttt{emcee} routine applied for wind models, where $M_{\rm Henv}$ is constrained by $M_{\rm ZAMS}$, the optimization converges at the lowest $M_{\rm ZAMS}$, minimizing the discrepancy between the inferred $M_{\rm Henv}$ (when treated as a free parameter) and the wind model prediction. If $M_{\rm ZAMS}$, rather than $M_{\rm Henv}$, is treated as a free parameter, applying wind models to SNe from massive, partially stripped RSGs will significantly underestimate their $M_{\rm ZAMS}$.

This issue has important implications for the inferred $M_{\rm ZAMS}$ distribution. In \citet{martinez22}, wind models are applied to fit the plateau light curves of a large sample of SNe II, yielding an $M_{\rm ZAMS}$ distribution that follows a power-law:
\begin{equation}
    \frac{dN}{dM_{\rm ZAMS}}\,\propto\,M_{\rm ZAMS}^{\,\Gamma_{M}},
\end{equation}
with upper- and lower-cutoffs at $M_{\rm up}$\,$\sim$\,21.5\,$M_{\rm \odot}$ and $M_{\rm low}$\,$\sim$\,9.3\,$M_{\rm \odot}$. The power-law index is estimated to be $\Gamma_{M}$\,=\,-6.35 to -4.07, which is significantly steeper than the Salpeter initial mass function (IMF; \citealt{imf}), where $\Gamma_{M}$\,=\,-2.35. This unexpectedly steep distribution is a common outcome when $M_{\rm ZAMS}$ is inferred from plateau light curve fits using wind models. For example, \citet{morozova18} enforced $\Gamma_{M}\,$=\,-2.35 but still found an unusual clustering of inferred $M_{\rm ZAMS}$ at the lower mass end (see their Figure 9). Similarly, \citet{silva24} reported an even steeper $\Gamma_{M}$, ranging from -11.65 to -4.13. However, such steep $M_{\rm ZAMS}$ distribution is not seen when other methods, including pre-SN images and nebular spectroscopy, are applied \citep{DB20,fang25c}.

The inconsistency between the observed $M_{\rm ZAMS}$ distribution and the Salpeter IMF is referred to as IMF incompatibility in \citet{martinez22}. Several possible modifications to pre-SN structure have been proposed to address this issue, including variations in the mixing length parameter (which affects RSG compactness) and enhanced mass loss. However, as discussed in \S4.3, modifying the mixing length parameter alone is unlikely to resolve this discrepancy, as it would result in excessively high $T_{\rm eff}$ values that are inconsistent with both field RSGs and the progenitor $T_{\rm eff}$ inferred from pre-SN imaging of SNe II.

In this section, taking SNe 2013by, 2013fs and 2014G as examples, we have demonstrated that, wind models will underestimate the $M_{\rm ZAMS}$ of SNe from partially stripped RSGs, mis-placing them to the low-end of $M_{\rm ZAMS}$ distribution. The unusually sharp $M_{\rm ZAMS}$ distribution inferred from wind models probably suggest that a large fraction of SNe II have lost more mass than predicted by standard stellar wind models, supporting the scenario of enhanced mass-loss as a key factor in shaping the diversity in SNe II light curves (\citealt{fang25a}).

Introducing partially stripped RSGs is crucial not only for constraining the mass-loss mechanism—the main focus of this work—but also for understanding the explosion mechanism of CCSNe. As shown in Table~\ref{tab:fit_zams}, when wind models are applied to SNe II originating from massive, partially stripped RSGs, $M_{\rm ZAMS}$ tends to be underestimated, while at the same time $E_{\rm K}$ is overestimated. For SN 2013by, wind models suggest it resulted from the energetic explosion ($E_{\rm K}\,\sim\,$1.5\,foe) of a relatively low mass RSG ($M_{\rm ZAMS}\,\sim\,11\,M_{\rm \odot}$). This scenario is inconsistent with the predictions of neutrino-driven explosion mechanism, as such high explosion energies are unlikely for low-mass progenitors (\citealt{stockinger20,burrows21,burrows24a,burrows24b,janka24}; though alternative mechanisms may be at play; see \citealt{soker24a,soker24b}).

\subsection{Can $M_{\rm ZAMS}$ be constrained from plateau light curve modeling?}
\begin{figure}
\epsscale{1.1}
\plotone{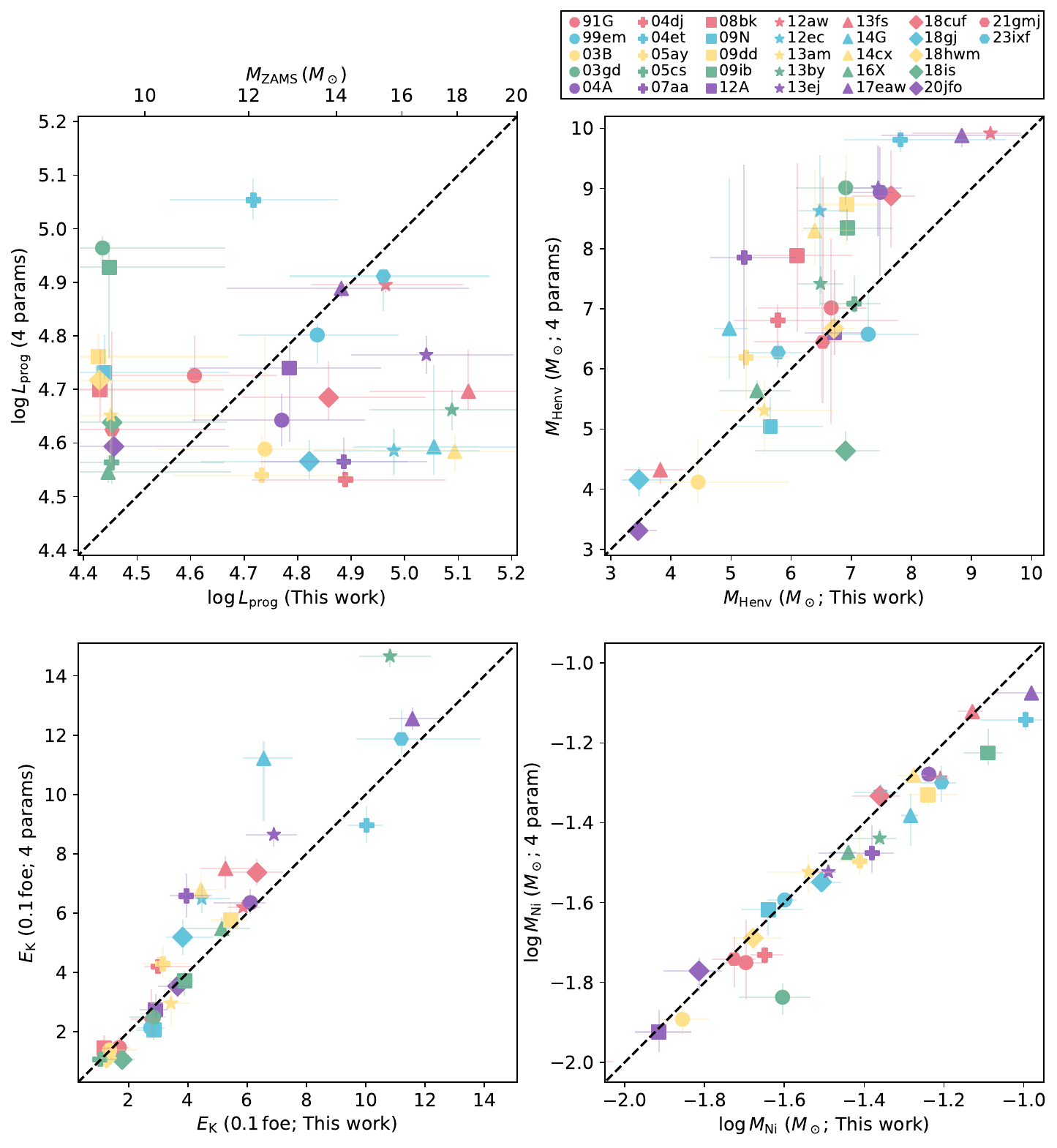}
\centering
\caption{Comparison of the measurements in this work and the inferred \{log\,$L_{\rm prog}$, $M_{\rm Henv}$, $E_{\rm K}$, log\,$M_{\rm Ni}$\} values from light curve modeling. Upper left panel: comparison between log\,$L_{\rm prog}$ inferred from nebular spectroscopy and the values from 4-parameter fit; Upper right panel: comparison between log\,$L_{\rm prog}$-weighted $M_{\rm Henv}$ and the values inferred from 4-parameter fit; Lower left panel: comparison between log\,$L_{\rm prog}$-weighted $E_{\rm K}$ and the values inferred from 4-parameter fit; Lower right panel: comparison between log\,$L_{\rm prog}$-weighted log\,$M_{\rm Ni}$ and the values inferred from 4-parameter fit. In all panels, different individual SNe are represented by markers and colors that are same as Figure~\ref{fig:main}. The black dashed lines indicate one-to-one correspondence.}
\label{fig:compare_4paras}
\end{figure}

In \citet{hiramatsu21}, a modified version of the method discussed in \S5.1 is introduced (see also \citealt{ravi24}), where \{$M_{\rm ZAMS}$, $M_{\rm Henv}$, $E_{\rm K}$, log\,$M_{\rm Ni}$\} are fitted simultaneously to model the light curves of 3 SNe II with short plateau duration. In this approach, since $M_{\rm Henv}$ is treated as a free parameter, $M_{\rm ZAMS}$ affects the light curve properties through $R_{\rm prog}$. Compared to the method used in this work, this modification introduces an additional free parameter $M_{\rm ZAMS}$ without adding new constraints, making the system underdetermined. However, it remains important to assess the reliability of the inferred \{$M_{\rm ZAMS}$, $M_{\rm Henv}$, $E_{\rm K}$, log\,$M_{\rm Ni}$\} from this method, as it can be readily applied to larger samples where key constraints in this work, such as nebular spectroscopy or pre-SN imaging, are unavailable.

Similarly to \S3.1, we add an additional parameter log\,$L_{\rm prog}$ (equivalent to $M_{\rm ZAMS}$) for light curve interpolation, and use the same \texttt{emcee} routine to model the observed light curves. The log\,$L_{\rm prog}$ inferred from this method is compared with those independently derived from nebular spectroscopy. The other 3 parameters, \{$M_{\rm Henv}$, $E_{\rm K}$, log\,$M_{\rm Ni}$\}, are also compared, with the labels 'This work' represent the log\,$L_{\rm prog}$-weighted values and '4 params' represent values inferred from simultaneously fitting the light curves with 4 parameters in this section, as shown in Figure~\ref{fig:compare_4paras}.

The first observation from this comparison is that log\,$L_{\rm prog}$, inferred via the 4-parameter fit in this section, exhibits significant scatter compared to log\,$L_{\rm prog}$ independently measured from nebular spectroscopy, and no correlation can be discerned. However, for the other three parameters, the values obtained from the two methods show good agreement, with standard deviations of 1.04\,$M_{\rm \odot}$ for $M_{\rm Henv}$, 0.11\,foe for $E_{\rm K}$ and 0.04\,dex for log\,$M_{\rm Ni}$. This is because the variations in these parameters are not very large between the models with log\,$L_{\rm prog}$ between 4.52 to 5.13 ($M_{\rm ZAMS}$ from 10 to 18\,$M_{\rm \odot}$). The weak constraint on log\,$L_{\rm prog}$, combined with the consistency in \{$M_{\rm Henv}$, $E_{\rm K}$, log\,$M_{\rm Ni}$\}, suggest that the properties of SNe II light curve are primarily determined by the latter three parameters, with log\,$L_{\rm prog}$ playing a relatively minor role. This behavior can be readily understood from scaling relations (e.g., \citealt{kasen09}): the plateau luminosity is predominantly determined by $E_{\rm K}$ and $R_{\rm prog}$, with a moderate dependence on $M_{\rm Henv}$. In our model grid, while $R_{\rm prog}$ varies only within a narrow range, $E_{\rm K}$ spans over an order of magnitude, making it the dominant factor controlling the luminosity. For the plateau duration, $E_{\rm K}$ and $M_{\rm Henv}$ are the primary regulators, with $R_{\rm prog}$ contributing only weakly. As a result, $E_{\rm K}$ essentially sets the plateau luminosity, and $M_{\rm Henv}$ is adjusted to match the duration, rendering $R_{\rm prog}$ (and thus log\,$L_{\rm prog}$, assuming fixed $T_{\rm eff}$) relatively unimportant and poorly constrained by light curve modeling alone.

\section{Conclusion}
In this work, we analyze the plateau phase light curves and nebular spectroscopy of 32 well-observed SNe II, aiming to bridge the gap between the surface and core of their RSG progenitors.

We begin by fitting the plateau light curve using a grid of models computed with \texttt{MESA}\,+\,\texttt{STELLA}. The progenitor models, with initial masses $M_{\rm ZAMS}$ ranging from 10 to 18\,$M_{\rm \odot}$, are evolved until core-carbon depletion using \texttt{MESA}. The hydrogen-rich envelope are artificailly stripped to varying degrees. For each SN II in the sample, we infer \{$M_{\rm Henv}$, $E_{\rm K}$, log\,$M_{\rm Ni}$\} at a fixed log\,$L_{\rm prog}$ (or equivalently, $M_{\rm ZAMS}$) by performing an MCMC fit to the multi-band light curves. Our results show that for any fixed log\,$L_{\rm prog}$, equally good fits to the observed light curves can be achieved, with the optimized \{$M_{\rm Henv}$, $E_{\rm K}$, log\,$M_{\rm Ni}$\} varying systematically with log\,$L_{\rm prog}$.

To incorporate nebular spectroscopy to break the above degeneracy, we convert the fractional flux of [O I] emission into log\,$L_{\rm prog}$ using an empirical mass-luminosity relation calibrated with a sub sample of SNe II with both nebular spectroscopy and pre-SN images available, following the method of \citet{fang25c}. This quantity is then used to weight the log\,$L_{\rm prog}$ dependent \{$M_{\rm Henv}$, $E_{\rm K}$, log\,$M_{\rm Ni}$\}. By comparing log\,$L_{\rm prog}$ with the log\,$L_{\rm prog}$-weighted $M_{\rm Henv}$ of the 32 SNe II in the sample, we find that almost all exhibit lower $M_{\rm Henv}$ than the prediction of the \texttt{KEPLER} models at a given log\,$L_{\rm prog}$. This suggests that the hydrogen-rich envelope is stripped more efficiently than expected from standard stellar wind models, and that this enhanced mass loss occurs across the entire log\,$L_{\rm prog}$ range. We further explore three possible scenarios that may account for the scatter in the log\,$L_{\rm prog}$-$M_{\rm Henv}$ relation: (1) alternative stellar wind; (2) binary evolution models; (3) more compact RSG progenitors. 

Although RSG models with smaller radii $R_{\rm prog}$, and hence higher $T_{\rm eff}$ at fixed log\,$L_{\rm prog}$, can reproduce the observed light curves without invoking enhanced mass loss, such a scenario would imply that roughly half of the SNe II originate from RSGs with $T_{\rm eff}\,>\,4300$\,K. Given the narrow and cooler $T_{\rm eff}$ distribution observed in field RSGs and inferred from pre-SN images, such a scenario appears less likely. Instead, variations in mass-loss rates at typical $T_{\rm eff}$ values around 3650\,K seem more plausible. In particular, binary interaction offers a compelling explanation for the non-monotonicity and large scatter in the log\,$L_{\rm prog}$–$M_{\rm Henv}$ relation. However, the high occurrence rate of partially-stripped RSGs cannot be accounted for by stable binary mass transfer alone without invoking fine-tuned initial orbital parameters distributions. This indicates that the mass-loss histories and the evolutionary pathways leading to SNe II, despite them being the most commonly observed class of CCSNe, are more complex than previously thought.

It is important to note that the analysis in this work is based on two major assumptions (1) one-dimensional stellar evolution and {shock propagation} models; (2) hydrostatic equilibrium. Relaxing one of these assumptions may result in significantly different envelope evolution compared to the models presented in this work (see \citealt{goldberg22a,ma24}). For example, pulsation may develop in massive stars, and depending on the phase of the pulsation at the onset of the explosion, the structure of the envelope may vary, which affects the plateau light curve (\citealt{goldberg_pulse}). Such processes may also induce dynamic mass loss, which could contribute to the formation of the large population of partially stripped RSGs reported in this work (\citealt{yoon_pulse}). In addition, our treatment of large-scale material mixing during shock propagation is deliberately simplified; a more sophisticated multidimensional approach might further revise the conclusions presented in this work \citep{goldberg22b,vartanyan25}.

Finally, we compare the parameters derived in this work with those obtained using other statistical methods. A commonly used approach for modeling the plateau light curves of SNe II is to treat $M_{\rm ZAMS}$ as a free parameter, and $M_{\rm Henv}$ is uniquely determined by $M_{\rm ZAMS}$ through stellar wind prescriptions (wind models). To assess the effect of this assumption, we apply the same \texttt{emcee} method to infer \{$M_{\rm ZAMS}$, $E_{\rm K}$, log\,$M_{\rm Ni}$\}, imposing the $M_{\rm ZAMS}$-$M_{\rm Henv}$ relation from \texttt{KEPLER} models, for three well-studied SNe: 2013by, 2013fs, and 2014G.  Our results show that, applying wind models to fit the plateau light curves of SNe II originated from massive, partially stripped RSGs  will significantly underestimate $M_{\rm ZAMS}$. This effect helps explain the unusually high fraction of SNe II with low $M_{\rm ZAMS}$ inferred from wind models (\citealt{morozova18,martinez22,silva24}).

Another commonly adopted method is to fit \{$M_{\rm ZAMS}$, $M_{\rm Henv}$, $E_{\rm K}$, log\,$M_{\rm Ni}$\} simultaneously. Applying this approach to the SNe II sample in this work, we find that the inferred $M_{\rm ZAMS}$, represented by log\,$L_{\rm prog}$, shows no correlation with log\,$L_{\rm prog}$ independently measured from nebular spectroscopy. However, \{$M_{\rm Henv}$, $E_{\rm K}$, log\,$M_{\rm Ni}$\} are consistent with their log\,$L_{\rm prog}$-weighted counterparts derived in this work. The inconsistency in log\,$L_{\rm prog}$, combined with the consistency in \{$M_{\rm Henv}$, $E_{\rm K}$, log\,$M_{\rm Ni}$\}, suggests that plateau light curve properties are primarily governed by the latter three parameters, while log\,$L_{\rm prog}$ plays a relatively minor role. This implies that the plateau light curve only contains information of the progenitor's surface, while the core properties, which are critical for understanding the explosion mechanism and the pre-SN evolutionary pathways, require additional constraints, such as those provided by pre-SN images or nebular spectroscopy.

\software{\texttt{MESA} \citep{paxton11, paxton13, paxton15, paxton18, paxton19, mesa23}; \texttt{STELLA} \citep{blinnikov98,blinnikov00,blinnikov06}; \texttt{SciPy} \citep{scipy}; \texttt{NumPy} \citep{numpy}; \texttt{Astropy} \citep{astropy13,astropy18}; \texttt{Matplotlib} \citep{matplotlib}; \texttt{emcee} \citep{emcee}}

\begin{acknowledgements}
The authors are grateful to Tatsuya Matsumoto and Ryosuke Hirai for insightful discussions, and the anonymous referee for comments that helped to improve the manuscript.
QF acknowledges support from the JSPS KAKENHI grant 24KF0080. TJM is supported by the Grants-in-Aid for Scientific Research of the Japan Society for the Promotion of Science (JP20H00174, JP21K13966, JP21H04997). KM acknowledges support from the JSPS KAKENHI grant JP20H00174, JP24H01810 and 24KK0070. SNe data used in this work are retrieved from the Open Supernova Catalog (\citealt{open_SN}), the Weizmann Interactive Supernova Data Repository (WISeREP; \citealt{wiserep}) and the UC Berkeley Filippenko Group's Supernova Database (SNDB; \citealt{sndb_origin,SNDB}).
\end{acknowledgements}

\begin{appendix}
The basic properties of the SNe II in this work are concluded
in Table~\ref{tab:sample_appendix}.

\setcounter{table}{0}
\renewcommand{\thetable}{A\arabic{table}}

\begin{deluxetable*}{cccccccc}
\centering
\label{tab:sample_appendix}
\tablehead{
\colhead{SN}&\colhead{explosion date}&\colhead{$\mu$}&\colhead{$E$\,($B\,-\,V$)}&\colhead{filters}&\colhead{log\,$L_{\rm prog}$}&\colhead{$M_{\rm ZAMS}$}&\colhead{References}
}
\startdata
SN1991G&48280&30.8 (0.07)&0.01&$VRI$&$4.62^{+0.16}_{-0.17}$&$10.97^{+1.67}_{-1.54}$&(1)\\
SN1999em&51476&29.4 (0.10)&0.10&$BVRI$&$4.83^{+0.15}_{-0.15}$&$13.18^{+2.18}_{-1.65}$&(2)(3)\\
SN2003B&52622&31.2 (0.25)&0.05&$BVRI$&$4.74^{+0.19}_{-0.23}$&$12.16^{+2.35}_{-2.25}$&(4)(5)(6)\\
SN2003gd&52717&29.3 (0.20)&0.14&$BVRI$&$4.45^{+0.23}_{-0.26}$&$<$\,10&(4)(7)\\
SN2004A&53010&31.5 (0.10)&0.21&$BVRI$&$4.78^{+0.16}_{-0.17}$&$12.69^{+2.03}_{-1.83}$&(8)(9)(10)\\
SN2004dj&53187&27.8 (0.10)&0.07&$BVRI$&$4.89^{+0.17}_{-0.18}$&$14.03^{+2.60}_{-2.14}$&(10)(11)\\
SN2004et&53270&28.9 (0.30)&0.36&$BVRI$&$4.71^{+0.16}_{-0.16}$&$11.89^{+1.83}_{-1.65}$&(4)(12)(13)\\
SN2005ay&53456&31.2 (0.10)&0.10&$BVRI$&$4.72^{+0.16}_{-0.15}$&$12.02^{+1.93}_{-1.52}$&(4)(14)(15)\\
SN2005cs&53549&29.3 (0.33)&0.05&$BVRI$&$4.42^{+0.25}_{-0.26}$&$<$\,10&(4)(16)(17)\\
SN2007aa&54131&32.1 (0.30)&0.02&$BV$&$4.88^{+0.15}_{-0.15}$&$13.87^{+2.34}_{-1.73}$&(18)(19)\\
SN2008bk&54550&27.7 (0.10)&0.02&$BVRI$&$4.45^{+0.22}_{-0.23}$&$<$\,10&(6)(20)\\
SN2009N&54848&31.7 (0.18)&0.13&$BgVRrIi$&$4.45^{+0.24}_{-0.27}$&$<$\,10&(21)\\
SN2009dd&54925&30.7 (0.15)&0.45&$BVRI$&$4.43^{+0.25}_{-0.23}$&$<$\,10&(22)\\
SN2009ib&55041&31.9 (0.15)&0.16&$BgVRrIi$&$4.45^{+0.23}_{-0.26}$&$<$\,10&(23)\\
SN2012A&55933&30.1 (0.15)&0.04&$BgVRrIi$&$4.78^{+0.18}_{-0.17}$&$12.70^{+2.30}_{-1.81}$&(24)\\
SN2012aw&56002&30.0 (0.03)&0.03&$BVRI$&$4.96^{+0.14}_{-0.14}$&$15.05^{+2.44}_{-1.89}$&(25)\\
SN2012ec&56143&31.2 (0.13)&0.14&$BgVRrIi$&$4.97^{+0.15}_{-0.13}$&$15.14^{+2.78}_{-1.75}$&(26)(27)\\
SN2013am&56371&30.5 (0.40)&0.65&$VRI$&$4.45^{+0.21}_{-0.25}$&$<$\,10&(10)(28)\\
SN2013by&56404&30.8 (0.15)&0.20&$BgVri$&$5.08^{+0.16}_{-0.17}$&$16.85^{+3.15}_{-2.58}$&(29)(30)\\
SN2013ej&56497&29.9 (0.12)&0.06&$BgVRrIi$&$5.03^{+0.17}_{-0.15}$&$16.22^{+3.23}_{-2.30}$&(10)(31)(32)\\
SN2013fs&56571&33.5 (0.15)&0.04&$BgVRrIi$&$5.11^{+0.19}_{-0.16}$&$17.72^{+3.16}_{-2.83}$&(33)(34)\\
SN2014G&56669&31.9 (0.20)&0.21&$BgVRrIi$&$5.04^{+0.15}_{-0.14}$&$16.28^{+3.04}_{-2.17}$&(34)(35)(36)\\
SN2014cx&56901&31.3 (0.40)&0.10&$BgVRrIi$&$5.10^{+0.14}_{-0.16}$&$17.34^{+2.72}_{-2.55}$&(37)(38)\\
SN2016X&57405&30.9 (0.40)&0.04&$BgVRrIi$&$4.43^{+0.23}_{-0.26}$&$<$\,10&(39)(40)\\
SN2017eaw&57886&29.2 (0.20)&0.41&$BgVRrIi$&$4.91^{+0.23}_{-0.25}$&$14.27^{+3.98}_{-2.92}$&(41)(42)\\
SN2018cuf&58292&33.1 (0.30)&0.14&$BgVri$&$4.86^{+0.17}_{-0.16}$&$13.55^{+2.50}_{-1.79}$&(43)\\
SN2018gj&58127&31.5 (0.15)&0.08&$BVRI$&$4.82^{+0.20}_{-0.21}$&$13.16^{+2.97}_{-2.28}$&(44)\\
SN2018is&58133&31.6 (0.19)&0.43&$BVgri$&$4.43^{+0.22}_{-0.26}$&$<$\,10&(45)\\
SN2018hwm&58425&33.6 (0.20)&0.03&$BgVRrIi$&$4.43^{+0.23}_{-0.26}$&$<$\,10&(46)\\
SN2020jfo&58973&30.8 (0.20)&0.08&$BgVRri$&$4.45^{+0.21}_{-0.25}$&$<$\,10&(47)(48)(49)(50)\\
SN2021gmj&59293&31.4 (0.20)&0.05&$BVRI$&$4.43^{+0.22}_{-0.22}$&$<$\,10&(51)(52)\\
SN2023ixf&60083&29.2 (0.02)&0.04&$BgVri$&$4.97^{+0.20}_{-0.19}$&$15.12^{+3.90}_{-2.51}$&(53)(54)\\
\enddata
\caption{SNe II sample in this work.}
\tablecomments{Columns: SN name, explosion date (MJD), distance modulus and uncertainty (mag), extinction, filters of the light curves, luminosity of the RSG progenitor from nebular spectroscopy (\citealt{fang25c}), ZAMS mass converted from the luminosity of the RSG progenitor with the mass-luminosity of the models in this work (in $M_{\rm \odot}$), references: (1)\,\citet{1991G}; (2)\,\citet{1999em}; (3)\,\citet{1999em2}; (4)\,\citet{faran14}; (5)\,\cite{anderson14}; (6)\,\cite{gutierrez17}; (7)\,\citet{2003gd}; (8)\,\cite{2004A1}; (9)\,\cite{2004A2}; (10)\,\cite{SNDB}; (11)\,\citet{2004dj}; (12)\,\cite{2004et1}; (13)\,\cite{2004et2}; (14)\,\cite{2005ay1}; (15)\,\cite{2005ay2}; (16)\,\cite{2005cs1}; (17)\,\cite{2005cs2}; (18)\,\cite{2007aa}; (19)\,\cite{maguire12}; (20)\,\citet{2008bk}; (21)\,\citet{2009N}; (22)\,\citet{inserra13}; (23)\,\citet{2009ib}; (24)\,\citet{2012A}; (25)\,\citet{2012aw}; (26)\,\citet{2012ec}; (27)\,\citet{jerk15}; (28)\,\citet{2013am}; (29)\,\citet{valenti_13by};(30)\,\citet{black17}; (31)\,\citet{2013ej}; (32)\,\citet{2013ej2}; (33)\,\citet{2013fs}; (34)\,\citet{valenti16}; (35)\,\citet{2014G2}; (36)\,\citet{2014G}; (37)\,\citet{2014cx1}; (38)\,\citet{2014cx2}; (39)\,\citet{2016X1}; (40)\,\citet{2016X2}; (41)\,\citet{2017eaw}; (42)\,\citet{2017eaw2}; (43)\,\citet{2018cuf}; (44)\,\citet{2018gj}; (45)\,\citet{2018is}; (46)\,\citet{2018hwm}; (47)\,\citet{2020jfo2}; (48)\,\citet{2020jfo}; (49)\,\citet{2020jfo3}; (50)\,\citet{kilpatrick23b};(51)\,\citet{2021gmj}; (52)\,\citet{murai24}; (53)\,\citet{singh24}; (54)\,\citet{2023ixf_nebular}.}
\end{deluxetable*}
\end{appendix}

{}
\end{document}